# Hierarchical Iterative Method in CFD Numerical Solution

Dehong Meng[a], Hao Yue[a], Hao Wang, Wei Li, Yuhang Qi, Rui Wang, Junwu Hong

Computational Aerodynamics Institute, China Aerodynamics Research and Development Center, Mianyang 621000, China

(a): These authors contributed equally to this work.

Corresponding author: Junwu Hong, hjw128@sina.com

Abstract: We propose a hierarchical asynchronous iterative method that differs from the traditional synchronous iterative method used across the entire flow field in conventional Computational Fluid Dynamics applications. This method forcibly divides the spatial region of the flow field into three layers: the boundary layer, the inner field, and the outer field. By adopting a novel approach of using different iteration steps for each layer, it significantly enhances computational efficiency. Using the hierarchical iterative method, numerical simulation studies were conducted on three typical benchmark models with different velocity ranges. Additionally, discussions were held regarding new modes such as using different control equations and computational parameters for each layer. The results based on structured grids indicate that, for the cases studied in this paper, the proposed method can achieve identical simulation results compared to traditional methods while only consuming 53.2% of the computational time of traditional methods, without significantly increasing manpower costs. This paper provides suggestions and discusses on the numerical applications of this novel iterative mode, and offers new insights for follow-up research based on this method.

## 1. Introduction

In traditional CFD (Computational Fluid Dynamics) numerical simulation methods, the iterative advancement of the flow field is carried out synchronously across the entire domain. In each iteration step, all spatial grid cells participate in solving the governing equations to achieve global exchange of flow field information and progressive advancement of flow parameters, ultimately obtaining the final converged flow field through repeated iterations. In this process, the number of iterations for all grid cells is completely consistent. Furthermore, in modern CFD simulations, the computational grid in the boundary layer has a very small scale in the normal direction to meet the high-resolution requirements in order to obtain high-precision boundary layer flow results. The aspect ratio of the first layer of grid cells can often exceed 10,000:1 and the corresponding volumes is $10^{13}$ to $10^{15}$ times smaller than that in the far-field. Since mainstream CFD methods utilize local time-stepping techniques that are highly dependent on the scale of local grid cells, the significant scale disparity causes the convergence speed of the boundary layer flow field to lag substantially behind that of the external flow field during time-marching iterations. Consequently, the boundary layer region becomes a convergence “bottleneck” for the entire flow field: in most cases, the external flow field remains in a state of “waiting” for the boundary layer to converge. Moreover, due to the delayed convergence of the boundary layer, numerical iterations in the external flow field effectively idle without progress, resulting in significant computational resource wastage.

A simple illustration of this situation is provided in Figure 1. The boundary layer flow field around the airfoil converges slowly, which can be supposed to a slow-moving turtle, while the outer flow field converges quickly, which can be supposed to a fast-running rabbit. We suppose that the distance a rabbit runs one step is equivalent to that of five steps for a turtle. Due to the interdependent convergence processes of the boundary layer and the external flow field—analogous to connecting them with a rope—the advancement of the faster-converging region (rabbit) cannot significantly surpass that of the slower-converging region (turtle). The ultimate goal of advancement is that turtle and rabbit arrive at the destination together. In traditional method, turtle and rabbit run the same number of steps and the remaining four steps for the rabbit become ineffective because of the rope. Comparing this illustration to the CFD iteration process, it is clearer that many of the outer flow field iterations, outside of the boundary layer, are essentially invalid. Meanwhile, these invalid steps can be eliminated entirely. Based on this analysis, this paper proposes an asynchronous iterative method that divides the flow field space into three layers. By modifying the traditional CFD iteration process, the flow field grids are forcibly stratified and subjected to different numbers of iterations within each layer. While ensuring that the numerical simulation results are completely consistent with the traditional approach, this method significantly reduces the computational resources required for flow field iterations.

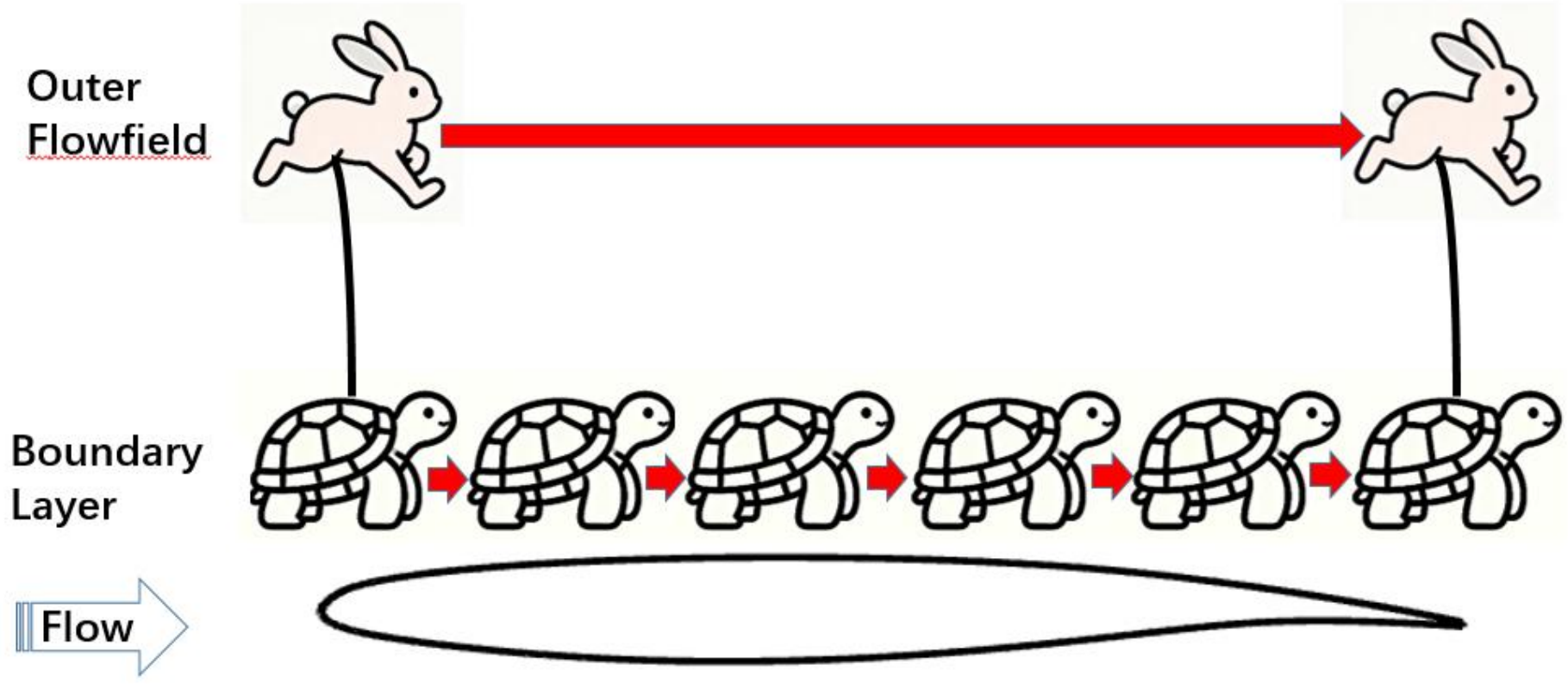

Figure 1 Principle Diagram of Hierarchical Iterative Method

Such a hierarchical iterative strategy raises a question: Could asynchronous iterations across different regions lead to desynchronized convergence processes in the flow field, potentially preventing overall convergence? In fact, we argues that due to the different convergence speeds in various regions of the flow field, even the traditional synchronous iteration over the entire flow field is inherently unsynchronized. During this iteration process, the outer flow field typically converges the fastest, followed by the inner flow field region (between the boundary layer and the outer flow), with the boundary layer converging last. Meanwhile, the phenomenon of unsynchronized convergence among different regions already exists in traditional methods. The number of iterations in faster-converging regions is reduced by introducing hierarchical asynchronous iteration and the unsynchronized behavior is alleviated. By the numerical result from multiple cases with different Mach numbers and angles of attack in this paper, it is shown that the flow field obtained by the asynchronous iterative mode are identical to those obtained by the traditional synchronous iterative mode. In most cases, the number of convergence steps is very similar between the two methods. In complex flow conditions, the asynchronous iterative mode can even converge faster than the traditional method.

The hierarchical iterative method proposed in this paper is fully based on the framework of traditional CFD methods and can be seamlessly integrated with CFD technical, such as structured/unstructured grid methods, multigrid techniques, and large-scale parallel computing methods. It is important to note that this paper introduces a new numerical iterative mode based on the physical essence of fluid flow, which is a strategy optimization within the original solution framework rather than a new solution algorithm. We also assume that readers are familiar with CFD algorithms and procedures, so detailed derivations of traditional CFD numerical methods will not be provided. The content of this paper primarily focuses on structured grid methods. Under the application of Geometric MultiGrid (GMG) techniques and parallel computing technologies, we present the strategy and complete implementation process of hierarchical iteration. Numerical validation for steady-state problems is conducted using multiple benchmark models with different speed range. This paper aims to inspire more CFD practitioners to explore the application based on this new iterative mode.

## 2. Hierarchical Iterative Method

For mainstream parallel CFD computations, this section will sequentially explain the hierarchical iterative method proposed in this paper by three stages: grid generation, parallel partitioning, and flow field solving. This method aims to divide the flow field into layers and conduct different iteration strategies in each layer, including varying numbers of iteration steps and different control equations, in order to maximize the conservation of unnecessary computational resources.

First, during the grid generation stage, the computational grid region is divided into three layers. The innermost layer is defined as the boundary layer, the middle region surrounding the boundary layer is defined as the inner field, and the outer region beyond the inner field is defined as the outer field. The schematic diagram of the region division is shown in Figure 2, which illustrates the region division for the DLR-F6 benchmark model: the yellow region represents the boundary layer, the green region represents the inner field, and the red region represents the outer field. The spatial boundaries of each region are also indicated in the figure. The theoretical basis for dividing the grid layers is as follows: the boundary layer flow is the most critical for flow field simulation and has the slowest convergence rate, thus it is designated as the “layer 1”; the inner field includes regions with significant gradients in flow parameters and covers areas where flow separation may occur, thus it is designated as the “layer 2”; the outer field includes regions with smaller changes in flow parameters, including the far-field region, and is designated as the “layer 3”. In this paper, the layer numbers (1, 2, and 3) in the figure are defined as the "layer sequence number."

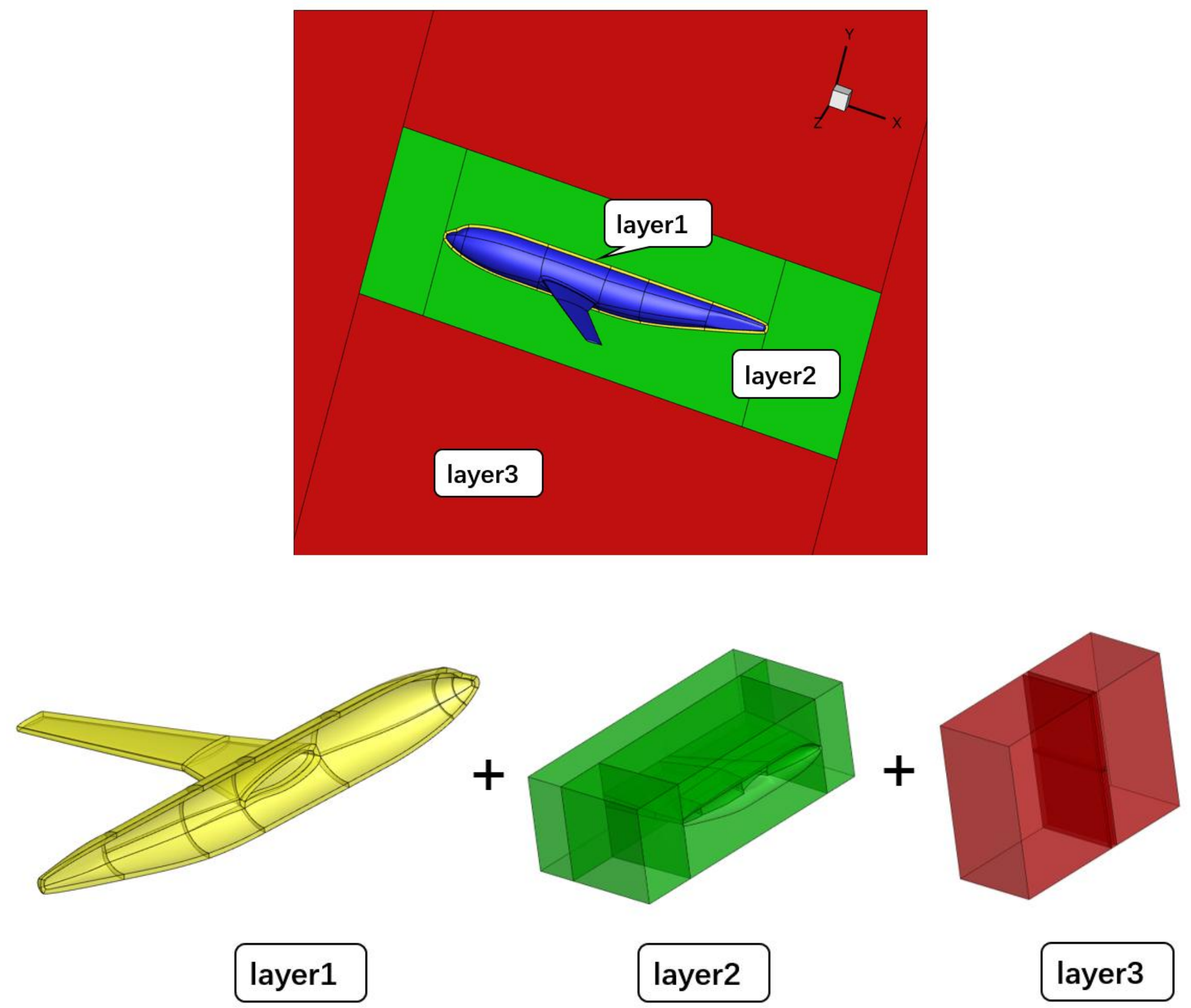

Figure 2 Schematic Diagram of Computational Region Division

Secondly, during the pre-processing stage of CFD parallel computation, it is necessary to partition the computational grid and distribute the partitioned grids to different processors for parallel solution. Under the iterative mode discussed in this paper, traditional parallel block partitioning strategies also need to be adjusted. If only serial computation is performed, this section can be skipped; however, current CFD simulations often cannot meet the computational demands imposed by large-scale grids using serial computation alone. Therefore, most practical CFD simulations are based on parallel computation. Currently, the most prevalent method for parallel computing in CFD is the "domain decomposition + message passing" strategy based on MPI (Message Passing Interface). For commonly homogeneous architecture cluster, all spatial grid cells should be partitioned and distributed as evenly as possible across all processors. Each processor exchanges flow parameter through the parallel communication surfaces (the surfaces of grid blocks that need to exchange flow field information with other processors) to achieve synchronized full-field computation. When a hierarchical iterative mode is introduced, the grid partitioning strategy needs to be adjusted. Traditional methods only ensure that the number of grid cells assigned to each processor is roughly the same, while the adjusted strategy ensures that the number of grid cells in each layer assigned to each processor is also roughly the same. For example, if there are 9,000,000 grid cells, with 3,000,000 each for the boundary layer, inner field, and outer field, and these are distributed among 20 processors, the traditional method would simply allocate about 450,000 grid cells per processor without specifying which specific region these cells come from. However, after adjusting the partitioning strategy, each processor must not only be allocated 450,000 grid cells, but also specifically be allocated 150,000 boundary layer grid cells, 150,000 inner field grid cells, and 150,000 outer field grid cells. Additionally, after parallel partitioning, each grid block must retain its "layer sequence number" information.

Finally, minor modifications are required for the flow solver code. In traditional structured-grid-based flow solvers, each iteration requires traversing all spatial grid blocks across the entire domain. In parallel computing, this process manifests as looping through all grid blocks localized to the current processor. After modification, the process of looping through all grid blocks on the current processor should be adjusted to independently loop through all layers on the current processor. This adjustment allows skipping the iteration processes for other layers on the same processor when only specific layer grids need to be iterated. Since " layer sequence number " have already been added to each grid block, this modification merely requires a simple adjustment to the original iteration control method, with minimal changes to the code. After the modification, the flow solver can independently adjust iteration parameters according to the user-selected iterative mode. For example, if the user specifies a “10-3-1” iteration template, meaning that within each cycle, the boundary layer grids iterate 10 times, the inner field grids iterate 3 times, and the outer field grids iterate once. Compared to the traditional “10-10-10” iteration (where all grid layers iterate 10 times), the “10-3-1” iteration reduces computational cost by approximately half. The exact reduction depends on the number of grid cells in each layer.

Specific implementation approaches are further provided for the strategic adjustments in the aforementioned three stages:

1) Grid Generation: During structured grid generation, to ensure the density and quality of the grid in the boundary layer region, the boundary layer grid is usually generated separately in blocks. This step is almost indispensable, especially in aerospace applications due to the importance of the boundary layer. Therefore, the Enforced Layering Strategy required in this paper will not significantly increase the work on grid generation. Users only need to add stratification processing for the inner field and outer field. Considering that the interface between these two layers is far from the wall, the grid topology is relatively simple, so adding stratification processing will not significantly increase the workload.

Additionally, after completing grid generation, each spatial grid block should be tagged with a "layer sequence number" to indicate its corresponding grid layer. The method used in this paper to add the "layer sequence number" information involves grouping the computational grid by "layer sequence number" in the grid generation software and sorting them in the order of boundary layer, inner field, and outer field. After completing grid generation, the block numbers at which each layer ends are output. For example, in the Trap wing high-lift configuration case used in this paper, the block numbers are 123-143-148, indicating that there are 148 total grid blocks, with blocks 1-123 being the “layer 1” (boundary layer), blocks 124-143 being the “layer 2” (inner field), and blocks 144-148 being the “layer 3” (outer field). However, arranging all the grids by "layer sequence number" is not mandatory; users only need to mark each grid block with some indication of its grid layer.

2) Parallel Partitioning: The implementation of hierarchical parallel partitioning varies depending on the user's partitioning strategy. In this paper, we briefly explain using the common greedy algorithm as an example. Other partitioning strategies can be analogously applied.

First, let's briefly introduce the greedy algorithm. The greedy algorithm aims to sort the grid blocks by the number of cells from largest to smallest and simultaneously sort the processors by their current computational capacity from largest to smallest, forming a grid list and a computational capacity list. The definition of current computational capacity is the remaining computational capacity after deducting the number of grid cells already allocated to the processor. For homogeneous architecture cluster, if no grid cells have been allocated to a processor, its current computational capacity is 100%. If it has been allocated an average number of grid cells, its current computational capacity is 0%.

During the grid allocation process, the greedy algorithm always attempts to place the first grid block in the grid list into the first processor in the computational capacity list and then makes a judgment. If the grid cells exceed the maximum number of grid cells the processor can accommodate, the grid is cut, and the remaining part of the grid is reinserted into the grid list according to the number of cells. Simultaneously, the processor is reinserted into the computational capacity list based on its remaining computational capacity. If the grid does not exceed the processor's capacity, the entire grid block is directly placed into the current processor, and the grid block is removed from the grid list. The current processor is then reinserted into the computational capacity list according to the remaining computational capacity. This process is repeated until the grid list is empty.

In order to implement hierarchical iterative parallel partitioning, the traditional strategy needs to be slightly adjusted. The rule for sorting the grid list by the number of cells is changed to a “dual-sorting” rule: First, ensure that in the grid list, all grid blocks with a smaller "layer sequence number" are always placed before those with a larger "layer sequence number," thus naturally forming the first level of sorting by "layer sequence number." Second, within grid blocks with the same "layer sequence number," sort them by the number of cells, thereby forming a “dual-sorting” mechanism prioritizing “layer sequence number” followed by "the number of cells". A comparison of the two strategies is shown in Figure 3. Following this sorting rule and using the greedy algorithm for allocation, the grid cells in “layer 1” will be evenly distributed to all processors first, followed by the distribution of the next layer. After cycling through all grid blocks in the grid list, the allocated processors will naturally form three layers of grid blocks, with each layer of grid cells being evenly distributed across all processors.

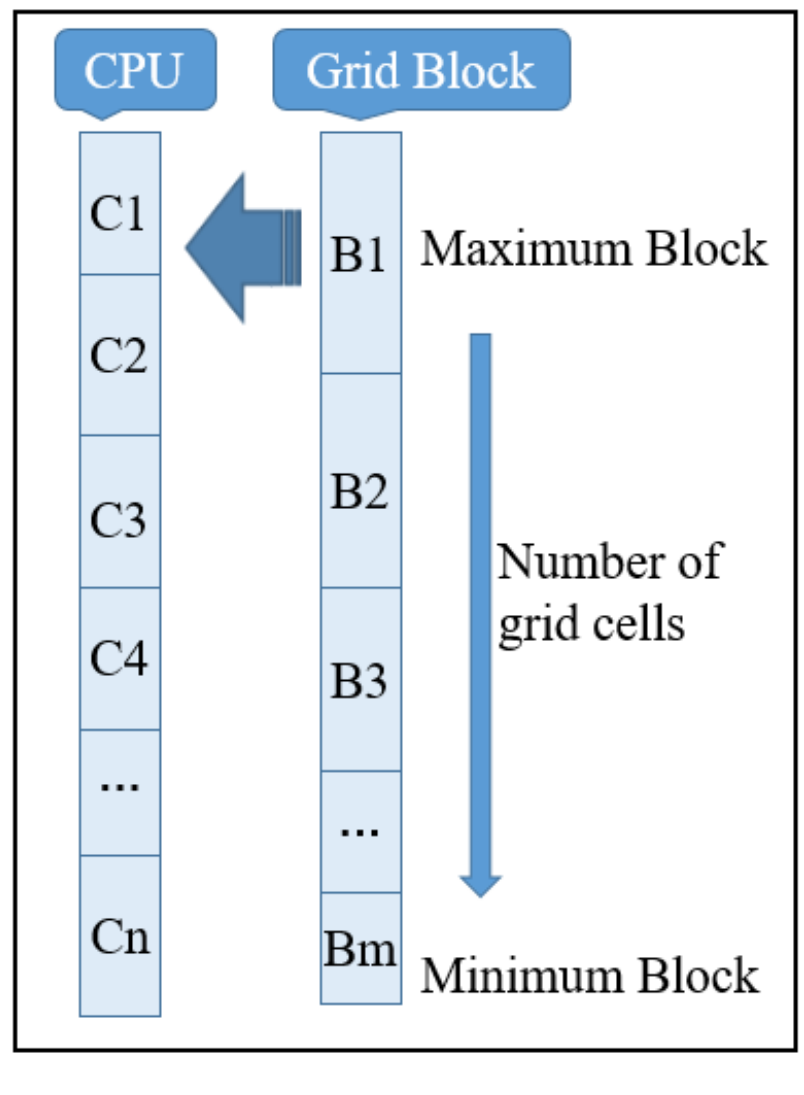

Original Strategy

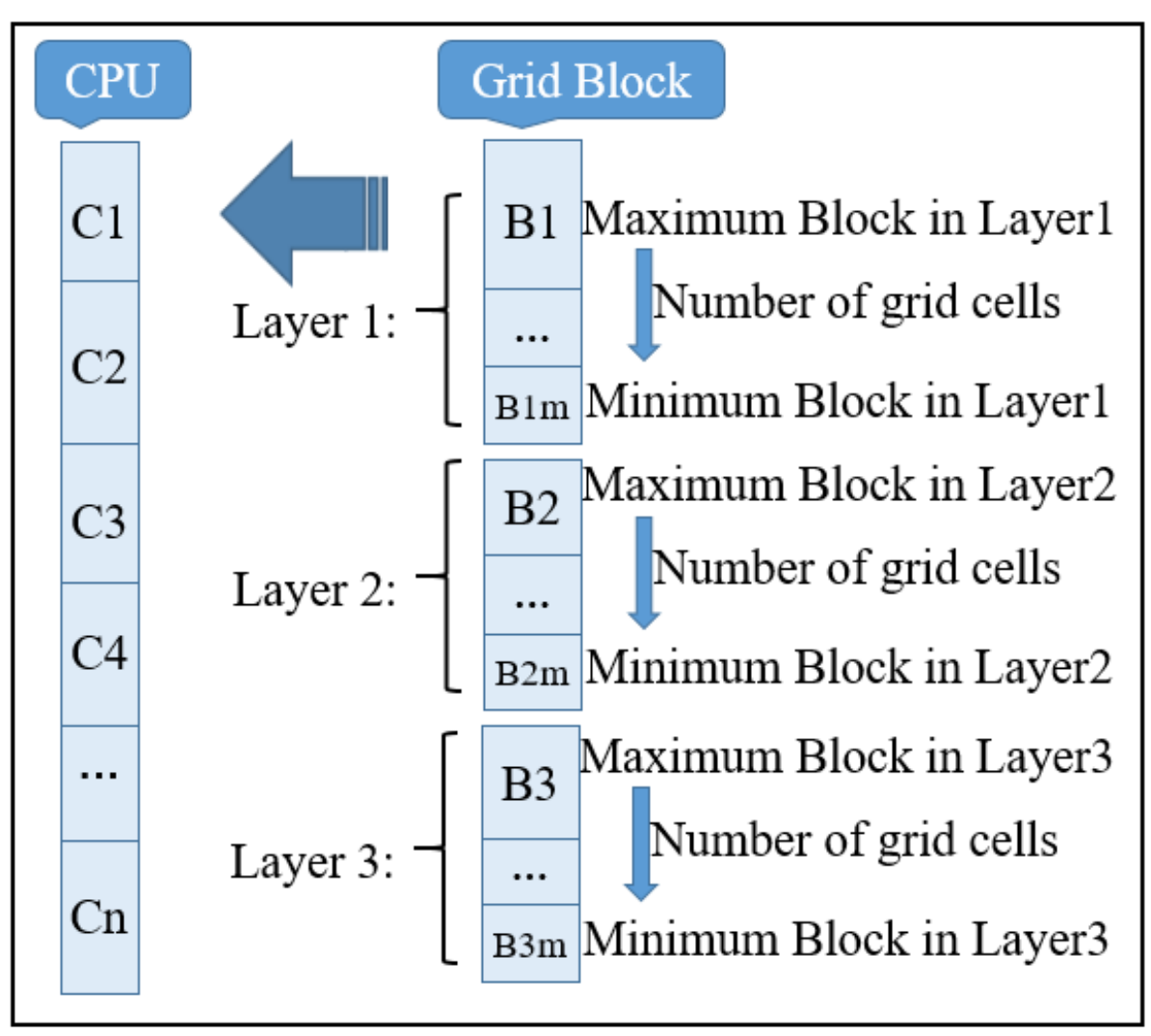

New Strategy

Figure 3 Comparison of the Original Strategy and the New Strategy for Greedy Algorithm Partitioning

3). Flow Field Solution: As mentioned earlier, adjusting the grid loop strategy in the solver requires minimal code changes. Most of the code modifications come from adjusting the parallel communication patterns. In traditional parallel communication, after each iteration step, all parallel communication interfaces need to exchange flow field information with their corresponding processors. However, under the new hierarchical iterative mode, taking the aforementioned "10-3-1" iterative mode as an example, as shown in Figure 4, the boundary layer（layer 1）flow field first undergoes 3 iterations. During the first and second steps of this process, the boundary layer flow field only exchanges information internally within the boundary layer, while the adjacent inner field does not need to exchange information. The reasons are:

(1). Since the inner field do not participate in the iteration in the next step, it is ineffective to transmit boundary layer information to the inner field.

(2). Because the inner field did not participate in the previous iteration step, its flow field parameters remain unchanged. Transmitting inner field information to the boundary layer would result in redundant information, which is also an ineffective operation. Therefore, this portion of the communication can be omitted.

Therefore, information exchange between the boundary layer and the inner field is only required after completing the third iteration step of “Layer 1” and before proceeding to “Layer 2” iterations. Additionally, during the computation of the boundary layer flow field, internal information exchanges in the inner field and outer field regions, as well as exchanges between the inner field and outer field, are clearly unnecessary and can also be omitted. This communication principle can be extended to other grid layers, ensuring that each layer only performs necessary communications after completing the current iteration, thereby saving communication time.

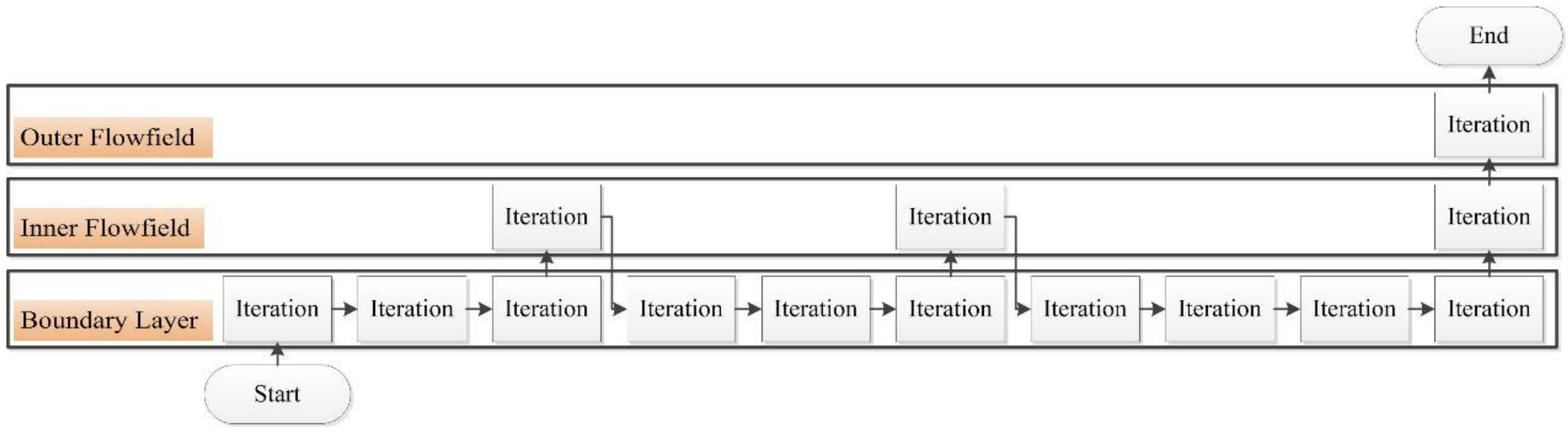

Figure 4 Iteration Process of the 10-3-1 Mode

Overall, the workload in the above three stages is relatively light. Specifically, the modification of parallel partitioning and the flow field solver are one-time efforts. For programmers familiar with the algorithm, depending on the structure of the solver used, the required code modifications amount to no more than 2-4 workdays. Although there is additional work in the grid generation process each time, it merely involves adding a hierarchical processing of the inner and outer fields, and the additional work can almost be ignored.

After completing the required code modifications and grid generation, the computational grid is segmented into three layers according to the flow field’s structural features. Each layer can be solved independently, with only a necessary information exchange needed before each iteration. Consequently, different iterative strategies can be employed for each layer to maximize computational efficiency.

Because boundary layer flows generally converge the slowest, internal flows converge next, and outer flows converge the fastest, it is shown that an iterative strategy should involve the most iterations in the boundary layer and the fewest in the outer field. In this paper, we combined hierarchical iterative method with multigrid techniques, using different solution templates based on a "10-3-1" mode, which is adjusted according to specific problems. The complete iteration process of the “10-3-1” template is as follows: Traditional multigrid methods perform 8-15 iterations on each grid level, while restricting and prolonging flow field variables between grids of varying densities to eliminate low-frequency errors in the flow field at the fastest possible rate. This study adopts a multigrid iteration cycle of 10 steps. On computational grids of varying densities, the traditional approach of performing 10 full-flow-field iterations per cycle is replaced with a Layering Strategy: 10 iterations for the boundary layer, 3 for the inner field, and 1 for the outer field, thereby reducing computational cost. In practice, the baseline “10-3-1” template is adjusted according to specific configurations and velocity ranges, as detailed in subsequent sections.

## 3. Computational Software and Numerical Methods

This work is conducted using the NNW-FSI software. The NNW (National Numerical Wind Tunnel) software platform was developed by the China Aerodynamics Research and Development Center (CARDC) in collaboration with over 30 renowned universities, more than 10 leading research institutes, and 3 high-tech software enterprises [1~4]. NNW-FSI is the FSI (Fluid-Structure Interaction) software within this suite. It primarily consists of four key modules: CFD solver, CSD (Computational Structural Dynamics) solver, coupling interface data transfer, and dynamic grid deformation. These modules are organized through a main control program using appropriate coupling methods to achieve numerical simulations of complex aircraft static and dynamic aeroelasticity problems. Given that other modules are less relevant to this paper, the following section focuses on the CFD solver module's computational methods.

The CFD module in the NNW-FSI software is based on structured grid technology and employs the finite volume method to discretize the Reynolds-Averaged Navier-Stokes (RANS) equations in arbitrary coordinates. Convection terms are discretized using the Jameson central difference scheme, MUSCL-Roe upwind scheme, VANLEER scheme, and AUSM series schemes. Diffusion terms are discretized using second-order central schemes. Turbulence models include the SA one-equation model and the SST two-equation model. The discretized equation system is solved using the LU-SGS method. Unsteady computations use the "dual-time stepping" approach, while low-speed computations employ preconditioning techniques. For complex geometries, static/dynamic overlapping grid methods or static/dynamic patched grid methods are utilized. Geometric multigrid technology and large-scale parallel computing are employed to accelerate convergence [5~9].

The governing equations are the conservative form of the Euler/NS equations in an arbitrary coordinate system $(\xi,\eta,\zeta)$, expressed as follows:

$$\frac{\partial \hat{Q}}{\partial \tau}+\frac{\partial \hat{E}}{\partial \xi}+\frac{\partial \hat{F}}{\partial \eta}+\frac{\partial \hat{G}}{\partial \zeta}=NVIS\cdot\left(\frac{\partial \hat{E}_v}{\partial \xi}+\frac{\partial \hat{F}_v}{\partial \eta}+\frac{\partial \hat{G}_v}{\partial \zeta}\right) \tag{1}$$

where:

$$\hat{Q}=Q/J$$

$$\hat{E}=(\xi_t Q+\xi_x E+\xi_y F+\xi_z G)/J \quad , \quad \hat{E}_v=(\xi_x E_v+\xi_y F_v+\xi_z G_v)/J$$

$$\hat{F}=(\eta_t Q+\eta_x E+\eta_y F+\eta_z G)/J \quad , \quad \hat{F}_v=(\eta_x E_v+\eta_y F_v+\eta_z G_v)/J$$

$$\hat{G}=(\zeta_t Q+\zeta_x E+\zeta_y F+\zeta_z G)/J \quad , \quad \hat{G}_v=(\zeta_x E_v+\zeta_y F_v+\zeta_z G_v)/J$$

J represents the Jacobian of the coordinate transformation, Q denotes the conservative variables in the Cartesian coordinate system, E, F, and G represent the convective fluxes, and Ev, Fv, and Gv represent the diffusive fluxes in the Cartesian coordinate system.

The finite volume method is applied to discretize Equation (1), leading to the following discrete equation set:

$$\frac{V}{\Delta\tau}\delta Q+\omega\left[\frac{\partial}{\partial\xi}(A\delta Q)+\frac{\partial}{\partial\eta}(B\delta Q)+\frac{\partial}{\partial\zeta}(C\delta Q)\right]=RHS \tag{2}$$

where, $RHS=-(\frac{\partial\hat{E}}{\partial\xi}+\frac{\partial\hat{F}}{\partial\eta}+\frac{\partial\hat{G}}{\partial\zeta})^{n}+NVIS\cdot(\frac{\partial\hat{E}_{v}}{\partial\xi}+\frac{\partial\hat{F}_{v}}{\partial\eta}+\frac{\partial\hat{G}_{v}}{\partial\zeta})^{n}$,

A, B, and C are the Jacobian matrices of the convective fluxes with respect to Q in the three directions. $\delta Q=Q^{n+1}-Q^{n}$ , $V=1/J$ . $\omega=1$ indicates first-order temporal accuracy, $\omega=1/2$ indicates second-order temporal accuracy. When NVIS = 0, it represents the Euler equations, and when NVIS = 1, it represents the Navier-Stokes equations.

In this study, the discretization of the convective terms is performed using the MUSCL-Roe scheme. The MUSCL-Roe scheme has low numerical viscosity and high numerical accuracy, making it suitable for detailed simulations of complex flows. It is one of the most widely used difference schemes in CFD numerical simulations today. In the validation calculations of this study, the SA one-equation model is used for low-speed problems, while the SST two-equation model is employed for transonic and supersonic problems. Considering the excellent acceleration effect of geometric multigrid techniques on structured grids, all case validations in this study utilize multigrid techniques, with a computational grid consisting of three levels of grids. To facilitate comparison, the number of multigrid inner cycles is set to 10 steps for all cases.

Additionally, the NNW-FSI software has conducted extensive research on parallel computing to address the increasing scale of computational grids. The software has undergone significant development and testing for ultra-large-scale cases (with over a billion grid cells), achieving excellent parallel computing performance. The NNW-FSI software has been applied to both large-scale homogeneous and heterogeneous cluster systems developed in China, enabling CFD numerical simulations of ultra-large-scale structured grids with 10 billions of grid cells on millions of cores. Specifically, using the MPI+OpenACC method on a heterogeneous cluster, the software simulated the aerodynamic elasticity benchmark AGARD 445.6 wing with 10.8 billion grid cells, achieving a parallel efficiency of 62.2% when scaling from 46,800 cores to 2,995,200 cores [10]. All the work presented in this paper is based on parallel computing.

## 4. Case Validation

To validate the methods described in this paper, three typical benchmark cases were tested. These cases are arranged in order of increasing complexity of their three-dimensional geometries: the supersonic flying wing benchmark CHN-F1, the transonic wing-body combination benchmark DLR-F6, and the low-speed high-lift benchmark Trap Wing full-span configuration.

For the aforementioned cases, this paper compares the computational accuracy and efficiency of traditional synchronous iterative methods with hierarchical asynchronous iterative methods. Additionally, the computational efficiency of different asynchronous iterative modes is compared.

The computational results are also compared with wind tunnel test data. In all examples, the sideslip angle is set to 0°, and only half of the model is computed.

When comparing CFD computational efficiency, it is important to note that some aerodynamic forces do not converge to fixed values. Therefore, establishing a universal convergence criterion is crucial. To ensure that different convergence behaviors can be precisely quantified in terms of the number of convergence steps, this paper adopts the following convergence criteria: using the final converged drag coefficient CD as the reference, an error range of ±0.0001CD is defined. The last iteration step where the drag convergence curve enters this convergence range is denoted as n. The iterations to convergence Nc is then defined as Nc = n×1.2. If the drag coefficient does not converge to a fixed value, its average value is used as the reference. The reason for choosing the drag coefficient as the convergence criterion is that the lift coefficient and moment coefficient can both be zero, making it difficult to quantify the convergence intervals for lift and moment. Figure 5 illustrates the convergence criteria. This paper does not discuss whether this criterion is sufficiently reasonable but aims to establish an objective criterion that can be precisely quantified and is fair and operational for different computational methods.

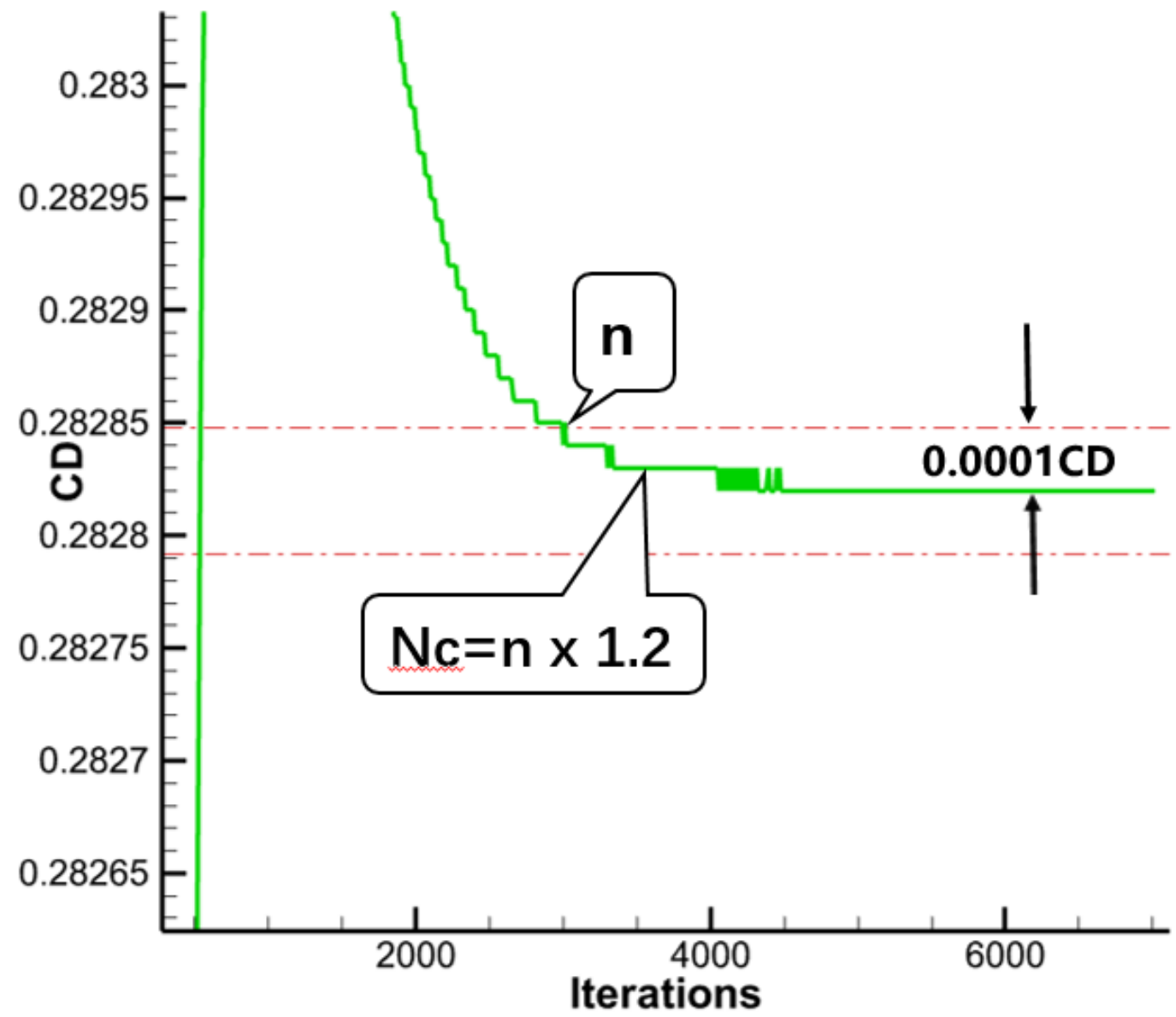

Figure 5 Schematic Diagram of Convergence Criteria

Additionally, after adopting a hierarchical asynchronous iterative mode, the definition of the number of iteration steps also needs to be changed: In traditional methods, one iteration step means that all spatial grid cells in the entire flow field participate in solving the governing equations once. In the hierarchical asynchronous mode, the number of iteration steps is determined by the number of times the boundary layer grids participate in solving the governing equations. Taking the previously mentioned "10-3-1" mode as an example, in one cycle of this mode, since the boundary layer grids are iterated 10 times, it is defined as 10 iteration steps. One iteration step in this mode can be understood as "the boundary layer grids iterate once, the inner field grids iterate 0.3 times, and the outer field grids iterate 0.1 times."

4.1 CHN-F1

CHN-F1 is a subsonic/transonic/supersonic flying wing benchmark case released by CARDC, primarily used to evaluate the aerodynamic simulation accuracy of CFD software for configurations with large sweep angles and small aspect ratios [11]. Table 1 provides the computational parameters for this case. This paper compares the traditional synchronous iterative method with the hierarchical iterative method for this configuration. The case covers a speed range from subsonic to supersonic, but this paper only compares the supersonic state.

Table 1 Computational Parameters for CHN-F1

| Mach number | 1.48 |
|---|---|
| Angle of attack (°) | -2~16 |
| sideslip angle (°) | 0 |
| Reynolds number | $1.10\times10^7$ |
| Reference area ($m^2$) | 0.117 |
| Reference length (m) | 0.5032 |
| Reference point for moments (m) | （0.3627, 0.0, 0.0） |

Figure 6 shows the surface grid of the computational configuration and the layered regions of each grid layer, with different areas distinguished by the aforementioned colors. The computational grid is divided into 48 blocks with a total of 6,559,744 grid cells. Blocks 1 to 29 are the boundary layers, containing 2,342,400 cells; Blocks 30 to 43 are the inner field, containing 2,198,016 cells; Blocks 44 to 48 are the outer field, containing 2,019,328 cells. The height of the first wall grid is 0.0012 mm, with $y^+ \approx 1$. Using the mean aerodynamic chord length L (the reference length in the table above) as the characteristic length, the turbulent boundary layer thickness is estimated to be 7.3 mm using &=0.37*L/$RE_L$**0.2. During the grid generation process, the grid height in the boundary layer region is set to approximately 14 mm, which is about twice the turbulent boundary layer thickness. The governing equations are the full Navier-Stokes equations, and the turbulence model used is the SST two-equation model.

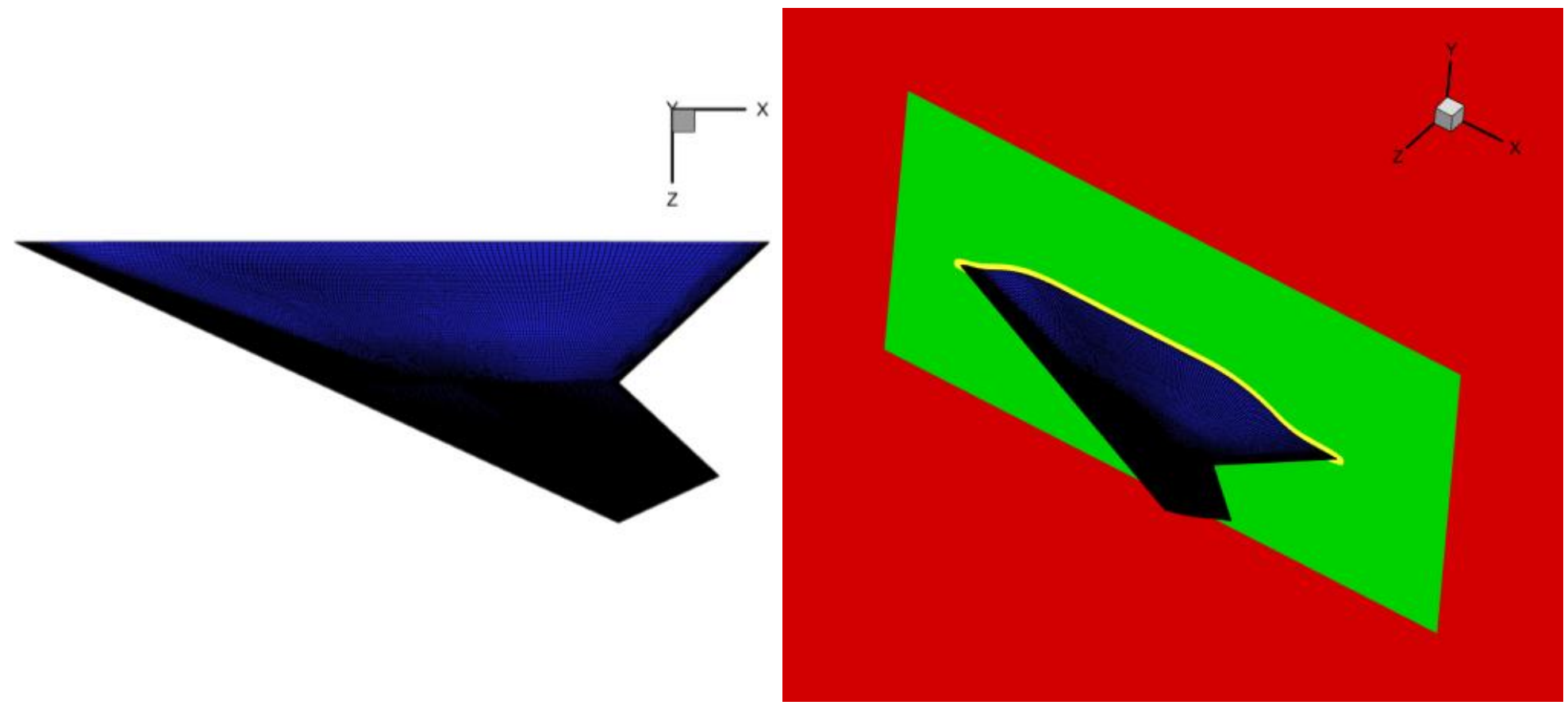

Figure 6 Computational Grid and Region Division for CHN-F1

In this paper, we first conducted parallel computations to compare the aerodynamic results under the new and old iterative modes. Since the grid partitioning scheme obtained through hierarchical splitting is just a special form of the original parallel splitting strategy, both iterative modes can use the same partitioning scheme for comparison. This study only compares the drag coefficient and lift coefficient. The configuration uses the "10-3-1" mode for iteration. Table 2 provides the differences in drag and lift coefficients obtained using traditional methods and the "10-3-1" mode for three typical angles of attack: negative angle of attack, moderate angle of attack, and high angle of attack. Within five significant digits, the results from both methods are identical. This indicates that, for such configurations at supersonic speeds, the hierarchical iterative method has equivalent numerical accuracy to the traditional method over a wide range of angles of attack. In the case of a 16° high angle of attack, where significant flow separation occurs on the upper surface, there is no difference in the aerodynamic results between the two modes.

Table 2 Comparison of Aerodynamic Forces Between New and Old Methods for CHN-F1

| Angle of attack(°) | method | CD | | CL | |
|---|---|---|---|---|---|
| -2 | Old | 0.025291 | 0.00% | -0.054665 | 0.00% |
| | 10-3-1 | 0.025291 | | -0.054665 | |
| 8 | Old | 0.066444 | 0.00% | 0.33068 | 0.00% |
| | 10-3-1 | 0.066444 | | 0.33068 | |
| 16 | Old | 0.19404 | 0.00% | 0.61754 | 0.00% |
| | 10-3-1 | 0.19404 | | 0.61754 | |

Table 3 presents the computational efficiency comparison between the new and old modes for three typical angles of attack. Here, "Old Iteration steps" refers to the number of convergence steps required by the traditional method, "Old Iteration Time" is the wall-clock time per step for the old method, and "Old Total Time" is the total wall-clock time required for convergence with the traditional method. "10-3-1 Iteration steps" denotes the number of convergence steps required by the "10-3-1" iterative mode, "10-3-1 Iteration Time" is the wall-clock time per step for the "10-3-1" iterative mode, and "10-3-1 Total Time" is the total wall-clock time required for convergence with the "10-3-1" mode. "Iteration Time Difference" represents the total convergence time difference between the two methods.

Table 3 Comparison of Computational Efficiency Between New and Old Methods for CHN-F1

| Angle of attack (°) | Old iteration Steps | Old Iteration time（s） | Old total time（s） | 10-3-1 Iteration Steps | 10-3-1 Iteration time（s） | 10-3-1 total time（s） | Iteration Time Difference |
|---|---|---|---|---|---|---|---|
| -2 | 1290 | 3.3067 | 4265.6 | 1290 | 1.7607 | 2271.2 | 53.25% |
| 8 | 2720 | | 8994.1 | 2780 | | 4894.6 | 54.42% |
| 16 | 1640 | | 5422.9 | 1680 | | 2957.9 | 54.54% |

Firstly, it can be observed that in the original mode, the single-step computation time for the three states is 3.3067 seconds. In the 10-3-1 hierarchical iterative mode, the theoretical single-step computational cost is 48.8% of the original method, but the actual average single-step computation time is 1.7607 seconds, which is 53.2% of the original method's computation time, resulting in a 4.4% discrepancy. This difference arises from several factors:

1). During the iteration process, there are some steps that cannot be accelerated, such as collecting flow field information, integrating aerodynamic forces, and data storage.

2). In the parallel partitioning phase, due to the hierarchical iteration requiring a three-layer partitioning, each layer has only about one-third of the total grid cells while also needing to satisfy the node constraints for three levels multigrid (the number of nodes in all directions of each grid block must be 4n+1). This makes uniform partitioning more difficult, leading to reduced uniformity in each layer's grid distribution, resulting in an efficiency loss of approximately 1% to 2%.

3). When iterating over a single layer of grids, the participating grid cells is only about one-third of the total grid cells. Although the amount of flow field information that needs to be communicated decreases, the reduction is not as significant as the decrease in grid cells, leading to an increased proportion of communication time. Additionally, in the hierarchical iterative mode, the number of communications increases compared to the original method. The original method communicates after completing iterations on all layers; for a 10-step cycle, this means communicating 10 times. In contrast, the hierarchical method requires communication after each layer's computation, resulting in 14 communications (10+3+1) in the 10-3-1 mode. Although the total communication volume is smaller than in the traditional method, the increased number of communications also results in some efficiency loss. The above numerical example was computed using 6 threads in parallel. To further investigate the adaptability of the new iterative mode to parallel computing and large-scale meshes, this paper refines the mesh of this example to 419,823,616 grid cells and partitions it for computation across 192, 384, and 768 cores. Table 4 presents a comparison of the single-step computational efficiency of the new and old methods under different thread counts. It can be observed that, under various thread counts, the difference in computational efficiency between the 10-3-1 pattern iteration time and the traditional method, compared with the theoretical computational workload, ranges from 5.1% to 7.2%. This difference is not significant when compared with the 6-thread results, and shows a trend of slow growth as the number of threads increases. These results fully demonstrate that the method proposed in this paper possesses good adaptability to large-scale meshes and multi-thread parallel computing.

Table 4 Computational Efficiency Comparison of Large-Scale Parallel Computing for the CHN-F1 Model

| Number of threads | Old Iteration time (s) | 10-3-1 Iteration time (s) | Iteration Time Difference |
| --- | --- | --- | --- |
| 192 | 7.0072 | 3.7783 | 53.92% |
| 384 | 3.6621 | 2.0497 | 55.97% |
| 768 | 1.8734 | 1.0446 | 55.76% |

Secondly, as shown in the table 3, the number of convergence steps between the new and old iterative modes is very close, with the new mode having an average increase of 1.8% in the number of convergence steps compared to the traditional method. This indicates that, for this supersonic case, the new mode does not significantly reduce the convergence rate despite reducing the number of inner and outer field iteration steps.

Figure 7 provides a comparison of the drag convergence curves under the new and old modes when $\alpha = 8°$. The "0.0001" dash-dot line represents the convergence interval for the drag coefficient, "old" represents the original method, and "10-3-1" and other labels represent different iterative modes. For comparison, the convergence curve for the "10-4-1" mode is also included, with the right side showing an enlarged view of the convergence curves. It can be observed that there are slight differences in the convergence curves of the three modes during the initial iterations, but they completely overlap at the end of the curves. The convergence curve for the "10-4-1" mode is closer to the old method, almost overlapping in the latter part of the curve, and enters the convergence interval slightly earlier than the "10-3-1" mode. However, since the single-step computation time for the "10-4-1" mode is longer than that of the "10-3-1" mode, the overall convergence efficiency is not as high as that of the "10-3-1" mode. Overall, in the three typical states, the new mode used only 54.07% of the computation time of the traditional method and achieved identical computational results.

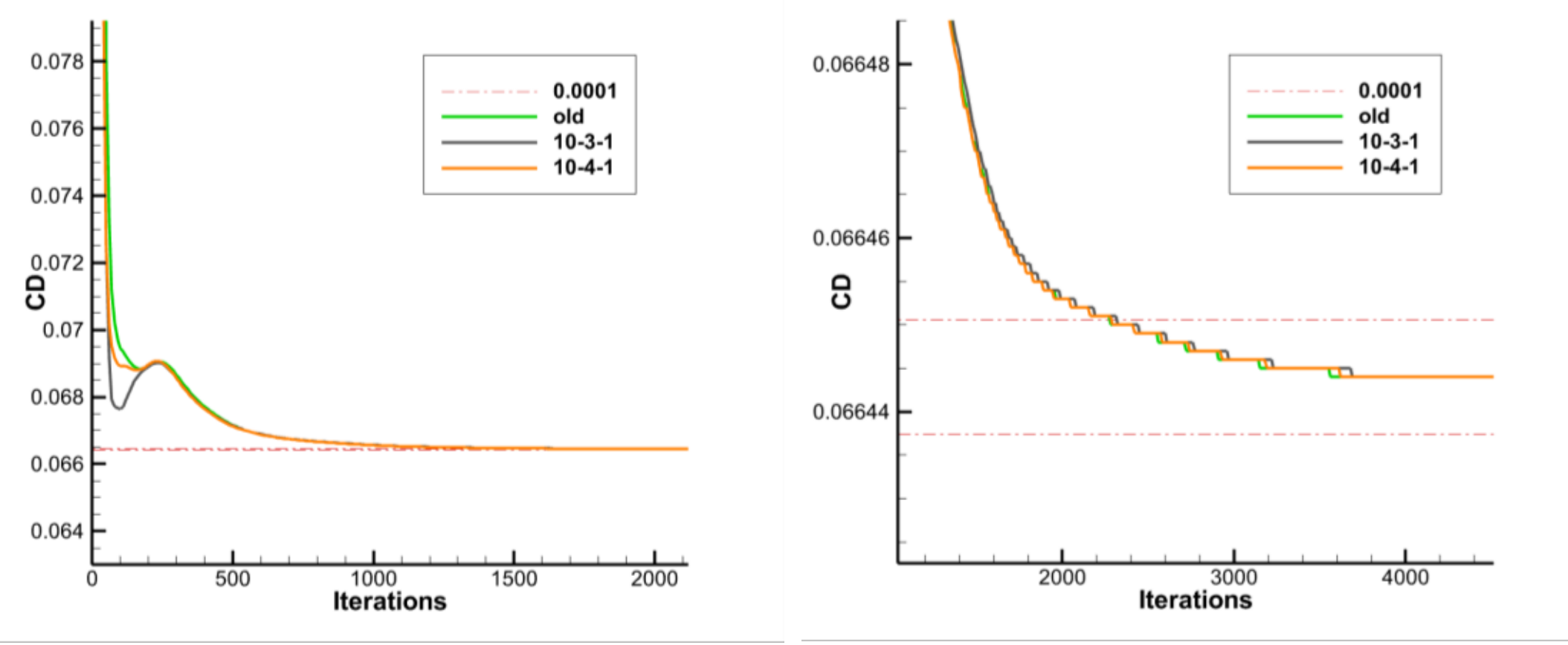

Figure 7 Comparison of Drag Coefficient Convergence Curves for Different Iterative Modes

Figure 8 shows the surface pressure contours and streamlines obtained using the new and old methods at α = 8°, with no visible differences.

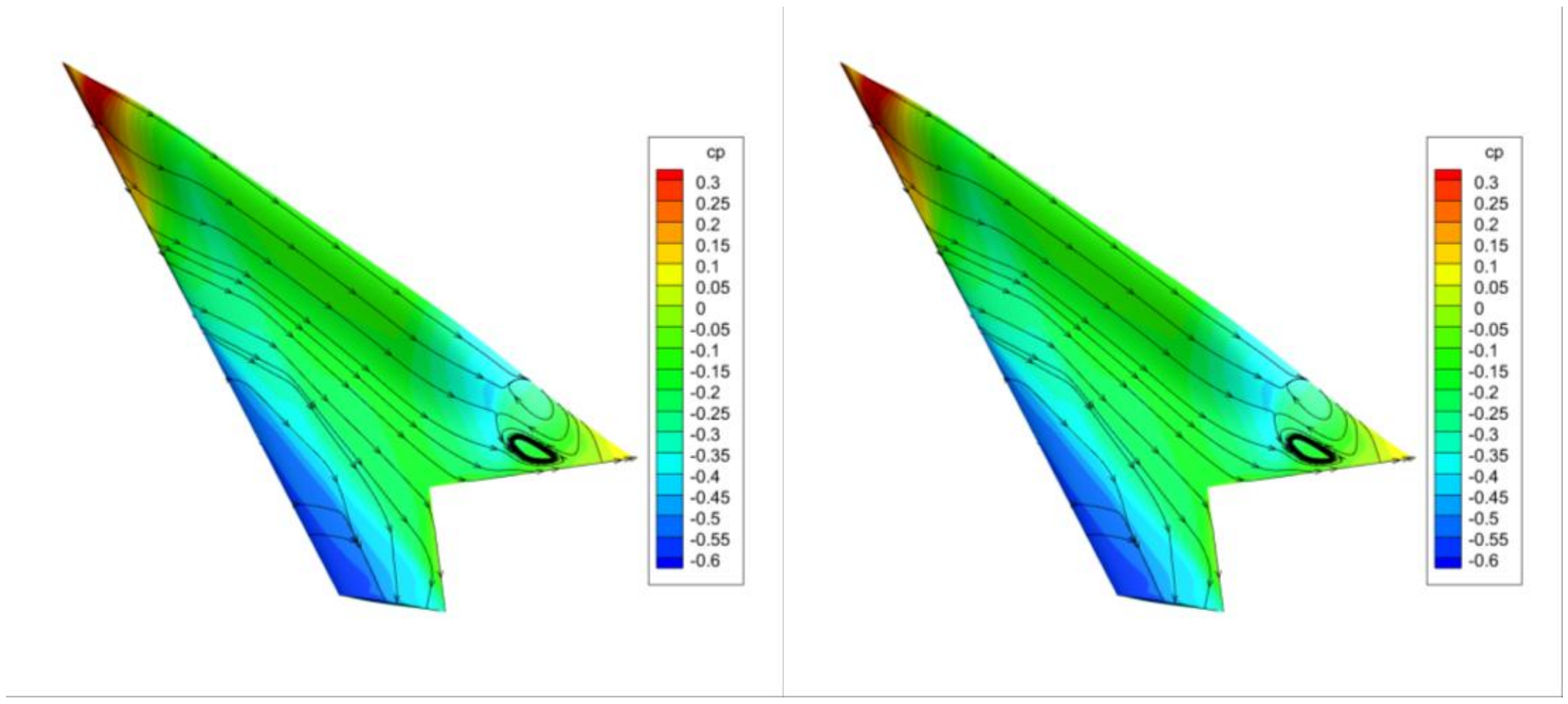

Figure 8 Surface Pressure Contours and Streamlines at AoA = 8° (Left: Original Method, Right: New Method)

Figure 9 compares the aerodynamic forces obtained using the "10-3-1" mode with the wind tunnel test results. It can be seen that, within the range from negative angle of attack to large angles of attack, except for the maximum angle of attack of 16°, the aerodynamic coefficients obtained by the method presented in this paper are in high agreement with the experimental values, which also demonstrates the reliability of the present method.

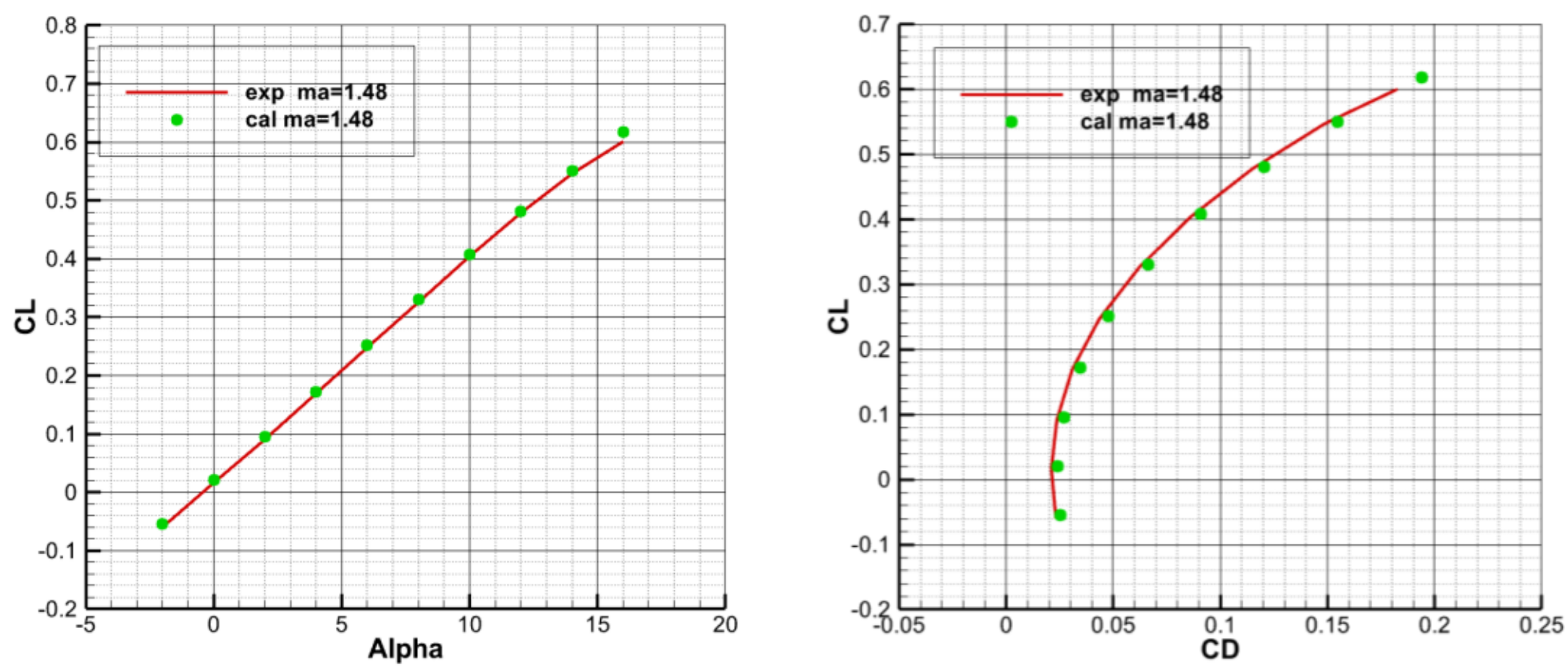

Figure 9 Comparison of Experimental and Computational Aerodynamic Coefficients

4.2 DLR-F6

The DLR-F6 configuration was released by the AIAA APATC (APplied Aerodynamics Technical Committee) at the Second Drag Prediction Workshop. This case is primarily used to evaluate the software's computational methods for predicting aerodynamic forces on transonic aircraft [12]. Table 5 provides the computational parameters for this case.

Table 5 Computational Parameters for DLR-F6

| Mach number | 0.75 |
| --- | --- |
| Angle of attack (°) | -3.0 ~1.5 |
| Sideslip angle (°) | 0 |
| Reynolds number | $3.0\times10^6$ |
| Reference area ($m^2$) | 0.1454 |
| Reference length (m) | 0.1412 |
| Reference point for moments (m) | （0.5049,0.0,-0.05141） |

Figure 10 shows the surface grid of the computational configuration and the layered regions of each grid layer. The computational grid is divided into 57 blocks with a total of 8,778,752 cells. Blocks 1 to 38 are the boundary layer with 2,606,080 cells; blocks 39 to 50 are the inner field with 3,493,888 cells; and blocks 51 to 57 are the outer field with 2,678,784 cells. The height of the first wall grid is 0.0012 mm, with $y^+\approx1$. The turbulent boundary layer thickness, estimated using the average aerodynamic chord length as the characteristic length, is 2.65 mm. During the grid generation process, the boundary layer region grid height was set to 9 to 11.5 mm (specific heights vary depending on different locations), which is about 3 to 4 times the actual boundary layer thickness. The governing equations are the full Navier-Stokes equations, and the turbulence model is the SST two-equation model.

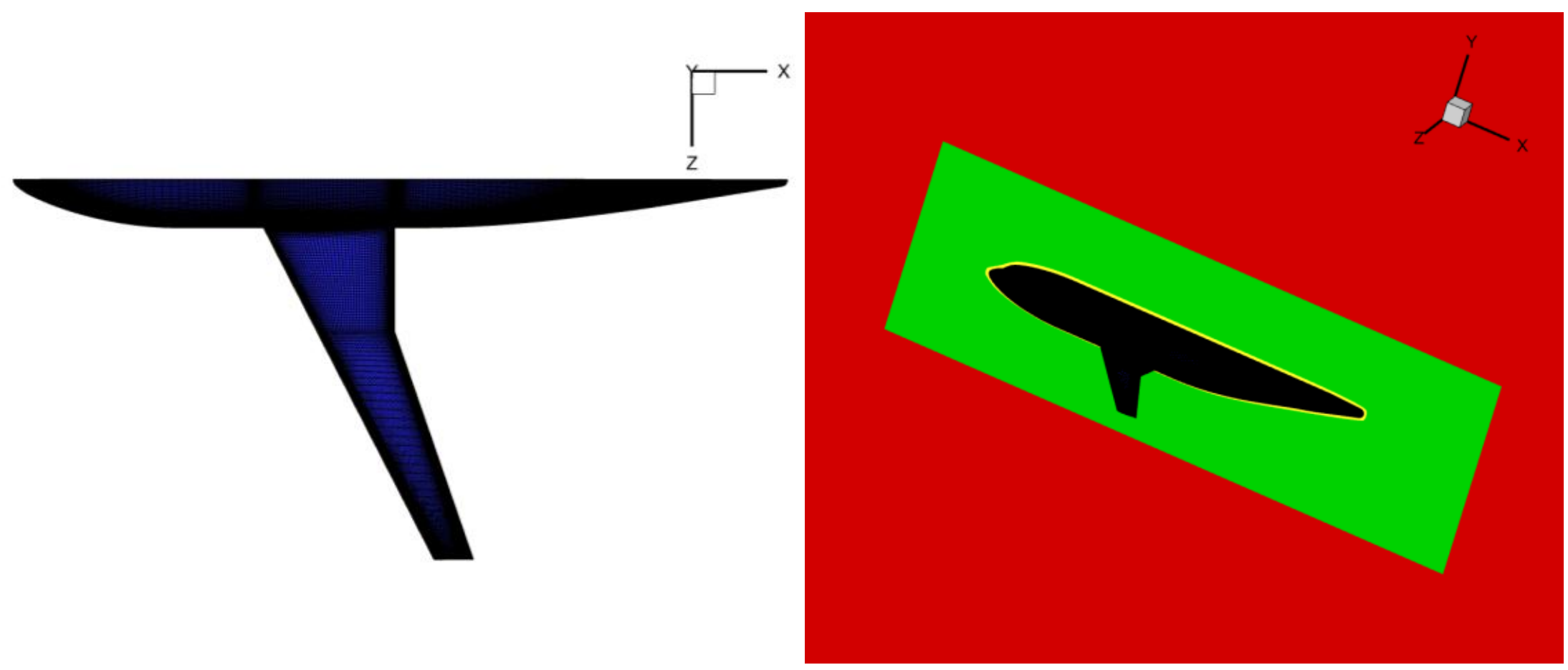

Figure 10 Computational Grid and Region Division for the DLR-F6

Unlike the previous examples, this configuration uses the "10-3-2" mode for iteration. Figure 11 shows the comparison of drag convergence curves for the original mode and different hierarchical iterative modes at a 1.5° angle of attack, with an enlarged view of the convergence curves on the right. It can be seen that the "10-3-2" mode and the "10-4-2" mode almost overlap with the traditional mode, while the "10-3-1" mode and the "10-4-1" mode highly overlap, both showing slight low-frequency oscillations at the end of the convergence curves. This analysis suggests that this phenomenon occurs due to insufficient iteration steps in the outer field, making it difficult to eliminate disturbances caused by low-frequency errors during the iteration process. The resulting disturbance waves oscillate within the flow field space. Adding one more iteration step in the outer field, changing "10-3-1" to "10-3-2," can resolve this issue. On the other hand, the "10-4-2" mode, which adds one more inner field iteration step compared to the "10-3-2" mode, has almost no impact on the convergence process, indicating that 3 inner field iterations are already sufficient. In this study, the "10-3-2" mode is used for all states of this configuration for comparison.

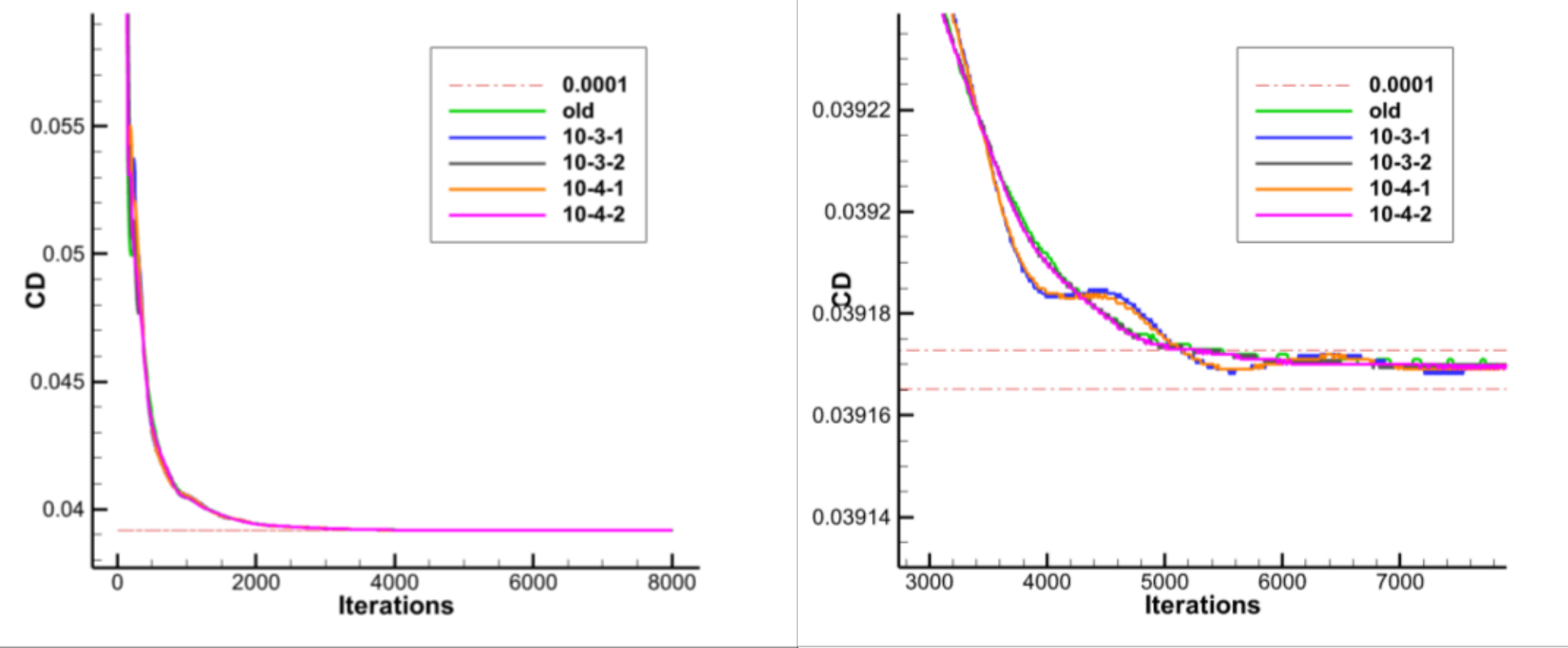

Figure 11 Comparison of Drag Coefficient Convergence Curves for Different Iterative Modes

Table 6 presents the differences in aerodynamic coefficients obtained using the traditional method and the "10-3-2" mode at three typical angles of attack. Within five significant digits, the results from both methods are almost identical. At $\alpha = 1.5°$, there is a very small difference in the last digit of the lift coefficient due to numerical iteration error. This indicates that the hierarchical iterative method achieves the same numerical accuracy as the traditional method for such configurations at transonic conditions.

Table 6 Comparison of Aerodynamic Forces Between New and Old Methods for DLR-F6

| Angle of attack(o) | method | CD | | CL | |
|---|---|---|---|---|---|
| -3 | Old | 0.021182 | 0.00% | 0.12828 | 0.00% |
| | 10-3-2 | 0.021182 | | 0.12828 | |
| 0 | Old | 0.029021 | 0.00% | 0.47560 | 0.00% |
| | 10-3-2 | 0.029021 | | 0.47560 | |
| 1.5 | Old | 0.039171 | 0.00% | 0.65028 | -0.0015% |
| | 10-3-2 | 0.039171 | | 0.65027 | |

Table 7 shows the comparison of computational efficiency between the two modes. As previously mentioned, there is a slight difference between the computation time of the "10-3-2" mode and its theoretical single-step computational cost, the reasons will not be reiterated here. In the new mode, except for the α = 0° state, the number of convergence steps for the other two states is slightly less than that of the original method. At α = 0°, the new mode's convergence curve oscillates around the drag coefficient values between 0.029024 and 0.029025 for nearly 400 steps before entering the convergence range, which does not satisfy the convergence criteria set in this paper. This results in an increase of 9.8% in the number of convergence steps compared to the original method. However, the actual convergence processes are quite similar. Overall, the average number of convergence steps for the three typical angles of attack is only about 1.6% higher than that of the original method. Additionally, the new method achieves the same computational results with only 54.67% of the computational time required by the original method.

Table 7 Comparison of Computational Efficiency Between New and Old Methods for DLR-F6

| Angle of attack(°) | Old iteration Steps | Old Iteration time（s） | Old total time（s） | 10-3-2 Iteration Steps | 10-3-2 Iteration time（s） | 10-3-2 total time（s） | Iteration Time Difference |
|---|---|---|---|---|---|---|---|
| -3 | 5550 | | 10091.7 | 5320 | | 5205.6 | 51.58% |
| 0 | 5010 | 1.8183 | 9109.8 | 5500 | 0.9785 | 5381.7 | 59.08% |
| 1.5 | 5910 | | 10746.3 | 5860 | | 5734.0 | 53.36% |

Figure 12 shows the surface pressure contours and streamlines obtained using the new and old methods at α = 1.5°, with no visible differences.

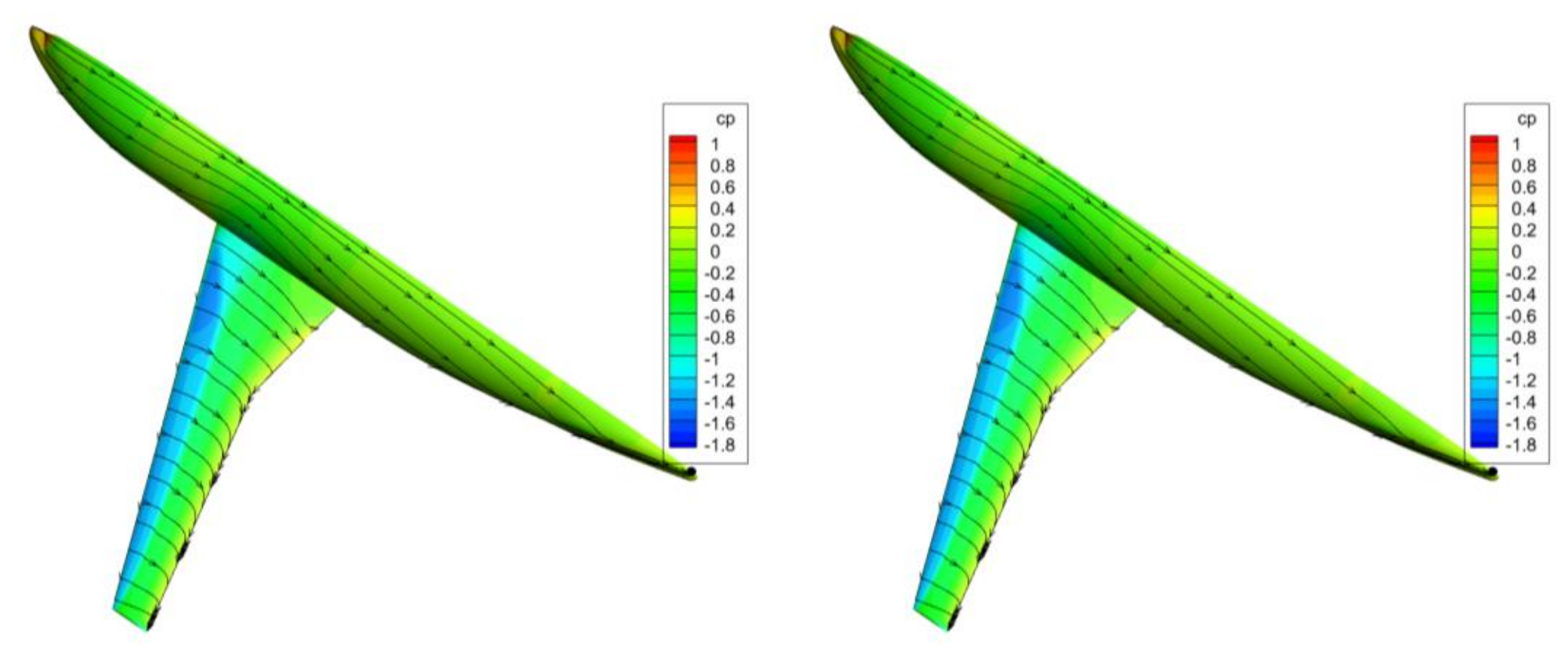

Figure 12 Surface Pressure Contours and Streamlines at AoA= 1.5° (Left: Original Method, Right: New Method)

Figure 13 compares the aerodynamic forces obtained using the "10-3-2" mode with wind tunnel test results. It can be seen that within the computational range, the lift coefficient obtained by the method in this paper is slightly higher than the experimental values, but the lift-to-drag ratio curve matches the experimental values very well.

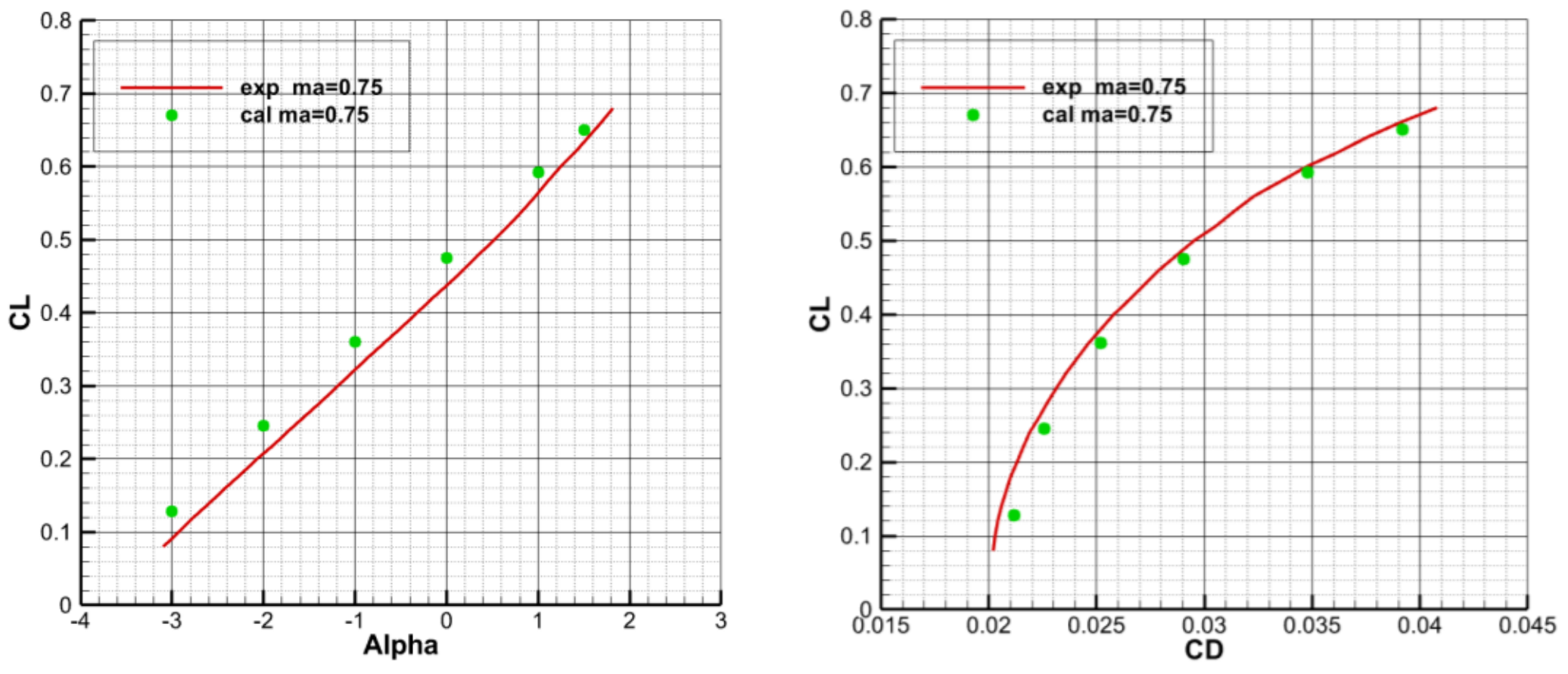

Figure 13 Comparison of Experimental and Computational Aerodynamic Coefficients

### 4.3 High-Lift Configuration Trap Wing

The flow field of high-lift configurations is generally closely associated with complex fluid physical phenomena such as flow separation, transition, and boundary layer mixing. For such complex flows with significant separation zones, accurately predicting the onset and development of separated flows is one of the challenges in CFD. The high-lift configuration tested in this study is the full-span Trap Wing configuration released during the 1st High-Lift Prediction Workshop (HiLiftPW-1) in 2010. This configuration is a three-segment design (slat/main wing/flap) with a long chord length and moderate aspect ratio, mounted on a simplified fuselage. The wind tunnel results used for comparison were obtained from tests conducted in 1999 at NASA Ames' 12-Foot Pressurized Wind Tunnel (PWT) [13]. The computational parameters for this case are provided in Table 8.

Table 8 Parameters for the Trap Wing

| Mach number | 0.15 |
|---|---|
| Angle of attack (°) | -3.45 ~24.4 |
| Sideslip angle (°) | 0 |
| Reynolds number | $1.5\times10^{7}$ |
| Reference area (m$^2$) | 2.046 |
| Reference length (m) | 1.006 |
| Reference point for moments (m) | （0.872,0.0,0.0） |

Figure 14 shows the surface grid of the computational configuration and the layered regions of each grid layer. The computational grid consists of 148 blocks with a total of 13,094,656 cells. Blocks 1 to 123 are the boundary layer with 4,375,168 cells; blocks 124 to 143 are the inner field with 4,196,352 cells; and blocks 144 to 148 are the outer field with 4,513,136 cells. The height of the first wall grid is 0.003 mm, with $y^{+}\approx1.5$. Based on the mean aerodynamic chord, the estimated height of the turbulent boundary layer is 13.7 mm. During the grid generation process, the boundary layer region grid height was set to 25 to 40 mm (specific heights vary depending on different locations), which is about 2 to 3 times the actual boundary layer thickness. The governing equations are the full Navier-Stokes equations and the turbulence model is Spalart-Allmaras. A low-speed preconditioning method was employed to accelerate convergence.

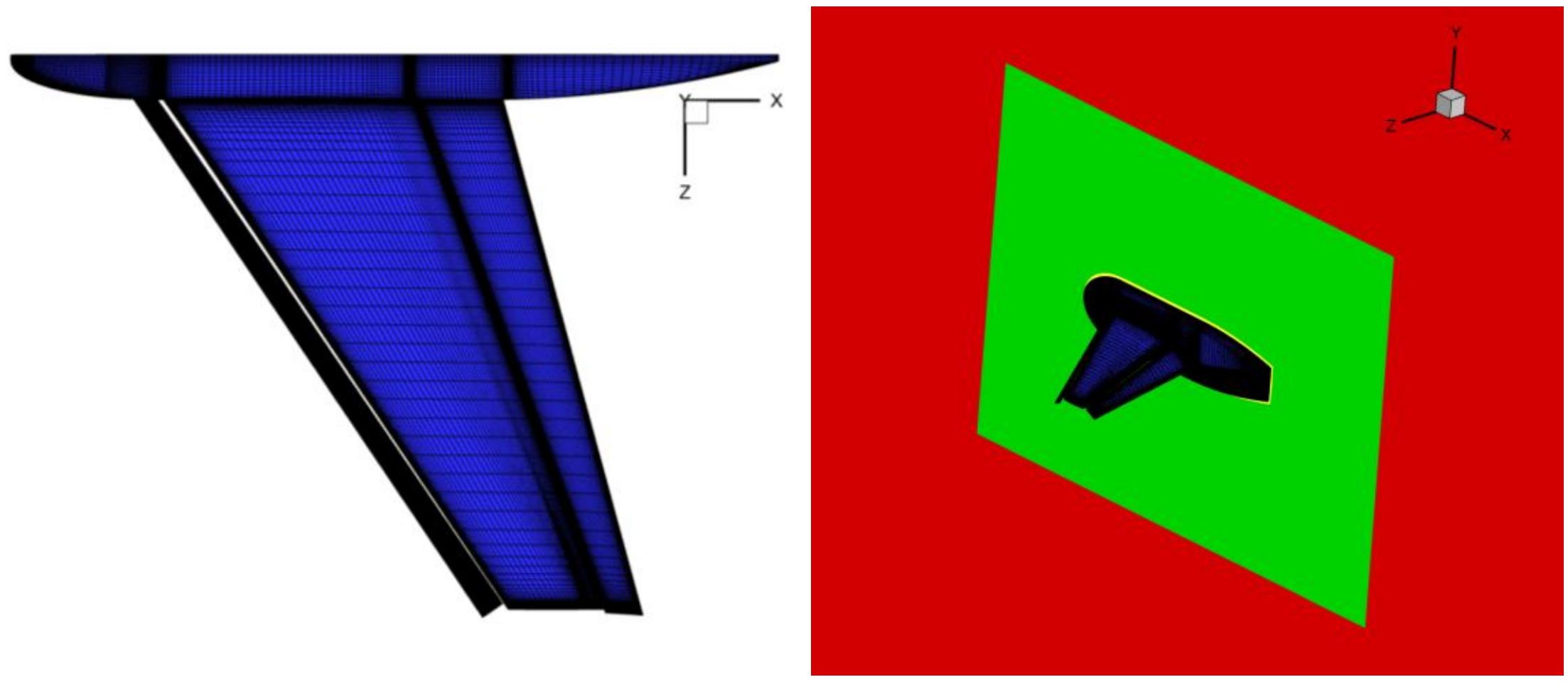

Figure 14 Computational Grid and Region Division for the Trap Wing

Compared to the two aforementioned test cases, the geometry of the Trap Wing is considerably more complex, resulting in higher complexity flow structure. The boundary layer mixing and flow separation between the three wing elements pose significant challenges for high-fidelity CFD simulations, while also providing an excellent benchmark for validating the applicability of the methodology proposed in this paper. Therefore, this test case is also a key focus of the present study. Moreover, it reveals significantly different phenomena compared to the aforementioned cases during testing, introducing new perspectives for the research method explored in this paper.

Table 9 presents the differences in the aerodynamic coefficients obtained by the new and the old methods at three typical angles of attack. Within five significant digits, the results are nearly identical. The only discrepancy is in the last digit of the lift coefficient at AoA = -3.45°, which is attributed to numerical iteration error. This indicates that the hierarchical iterative method achieves the same numerical accuracy as the traditional method for such complex configurations at low-speed conditions.

Table 9 Comparison of Aerodynamic Forces Between the New and Old Methods for the Trap Wing

| Angle of attack(°) | method | CD | | CL | |
|---|---|---|---|---|---|
| -3.45 | Old | 0.098086 | 0.00% | 0.39007 | -0.0026% |
| | 10-4-1 | 0.098086 | | 0.39006 | |
| 11.02 | Old | 0.28290 | 0.00% | 1.8791 | 0.00% |
| | 10-4-1 | 0.28290 | | 1.8791 | |
| 24.40 | Old | 0.58479 | 0.00% | 2.7308 | 0.00% |
| | 10-4-1 | 0.58479 | | 2.7308 | |

Figure 15 shows a comparison of convergence histories for different computational modes at AoA = -3.45°, with an enlarged view provided on the right. As can be seen, the number of iterations required for convergence is nearly identical between the old method and the new method with different modes. Increasing the number of inner iterations in the new method did not lead to accelerated convergence. It should be noted that under this condition, extensive flow separation occurs on both the upper and lower surfaces of the wing, particularly on the lower surface (see Figure 16). The instability of the separated flow results in exceptionally slow convergence of the flow field. Even after 30,000 iterations, the drag coefficient continues to decrease gradually. Nevertheless, this case once again demonstrates that even for conditions with large separation regions, the computational results and convergence behavior remain fully consistent between the old and new methods.

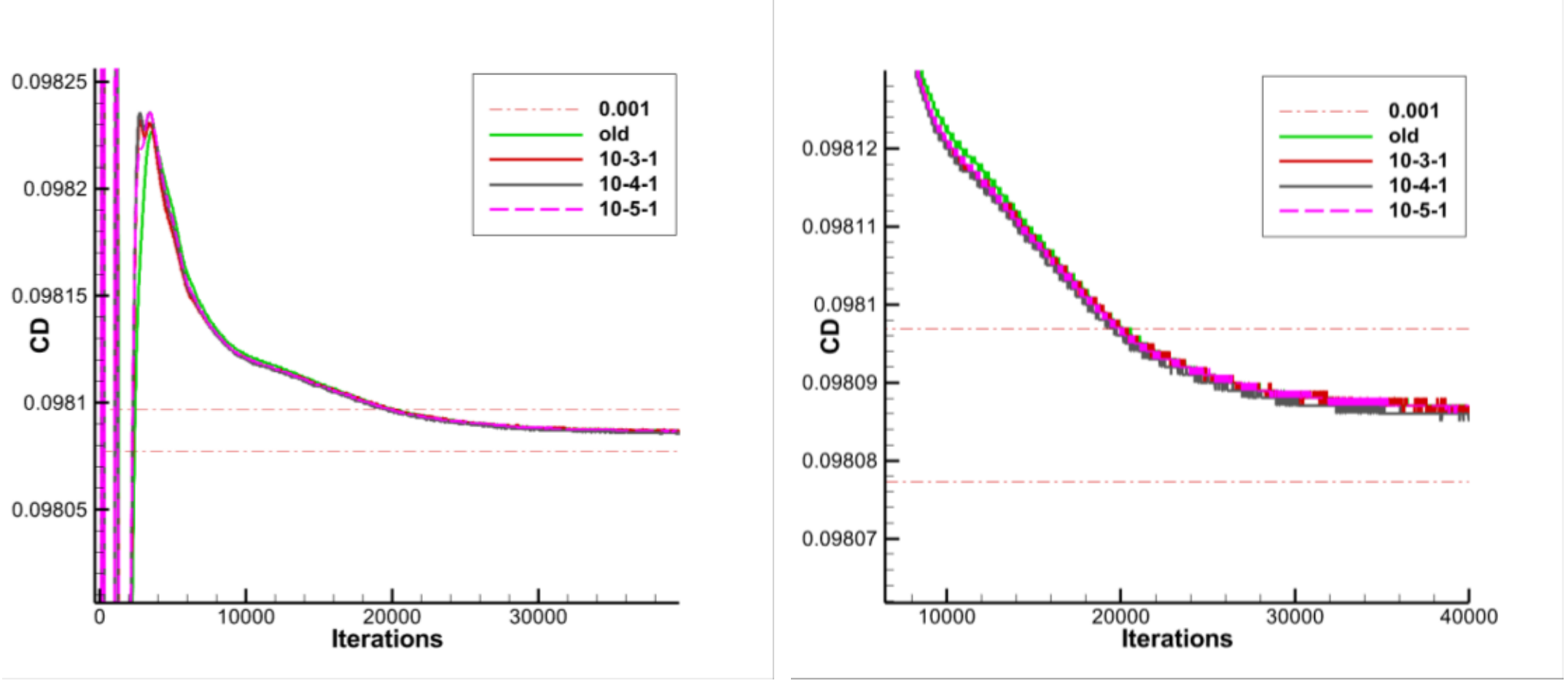

Figure 15 Comparison of Drag Coefficient Convergence Curves for Different Iterative Modes

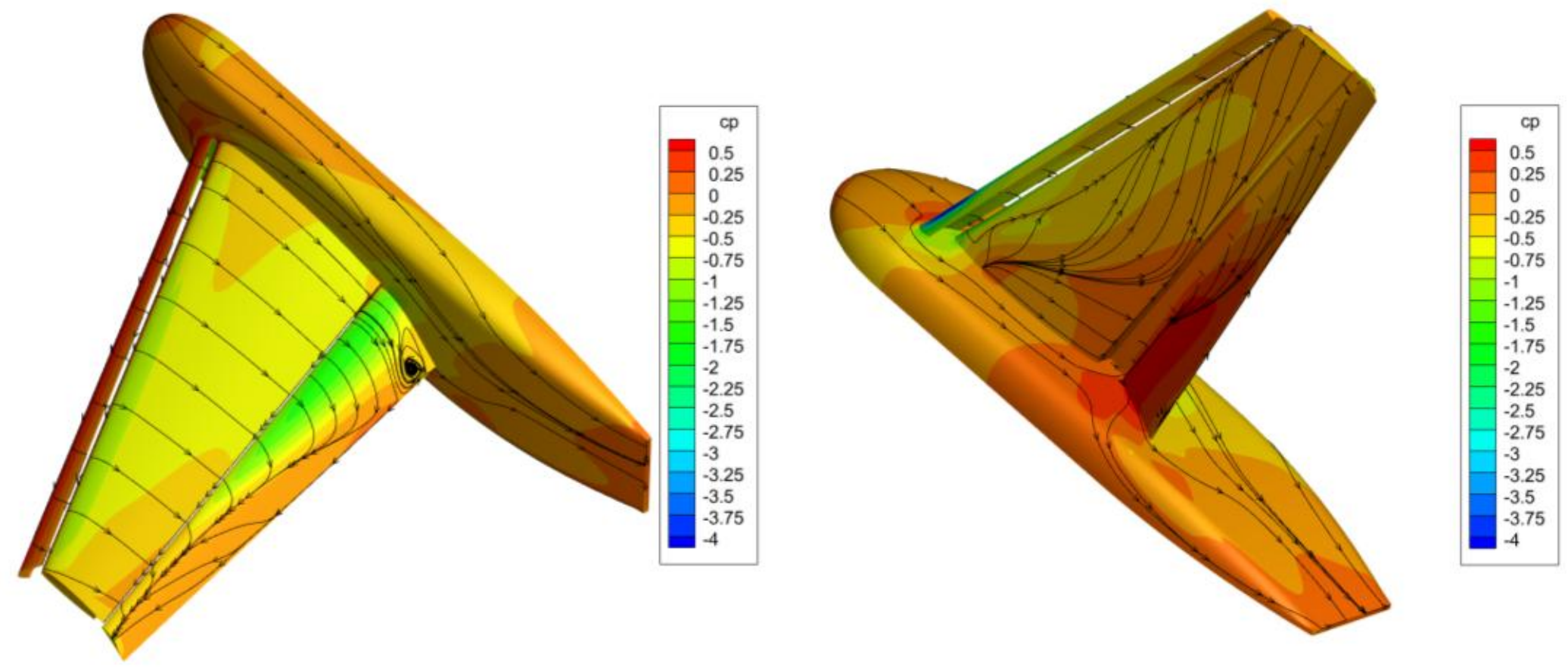

Figure 16 Pressure contours and streamlines on the upper and lower surfaces at AoA= -3.45°

Figure 17 compares the convergence histories of different computational modes at AoA = 11.02°, with a local magnification provided on the right. It can be observed that the convergence speed of the new mode “10-3-1” is significantly faster than that of the original mode that a phenomenon is not encountered in previous cases. While the original mode required 3,720 iterations to converge, the “10-3-1” mode achieved convergence in only 2,480 iterations, demonstrating a clear advantage over the original mode. This study suggests that under the new mode, the convergence behavior across different flow regions is more balanced, which may lead to an overall convergence rate that is not necessarily slower—and can even be faster—than that of the original method. The flow fields in the two previously mentioned cases exhibit simple structures, lacking the boundary layer mixing between the three wing segments present in this case, as well as the mutual interference among flow phenomena in different regions. Consequently, they do not demonstrate the accelerated convergence observed with the new mode in the present case. In the original mode of this test case, the higher number of inner field iterations leads to a mismatch in convergence speed between the inner field and the boundary layer. This results in an “overshoot” of flow parameters in the inner field region, causing the convergence speed to become slower than that of the new mode, which achieves a more balanced convergence across all regions.The convergence curve further shows that the traditional mode exhibits a higher spike around 600 iterations than the “10-3-1” mode, requiring more additional steps to pull back toward the converged values. It is noteworthy that the “10-3-1” mode also converges faster than both the “10-4-1” and “10-5-1” modes. Among three new modes, those with more inner iterations produce higher overshoots in the convergence curve before entering the converged region, thereby slowing down convergence rate. This further supports the observation that excessive inner iterations may disrupt the balance of convergence across the flow field and can even delay global convergence in such conditions.

As we use the rabbit and turtle analogy from Figure 1 to explain this phenomenon, it can be argued that due to the rabbit’s excessive speed, it moves far ahead of the turtle and then exerts a pull through the rope connecting them, interfering with the turtle’s movement. When the turtle’s path is unobstructed and relatively straightforward (analogous to a simple flow structure), this interference has little effect on its trajectory. However, when obstacles are present and multiple route choices exist along the turtle’s path (similar to the complex flow condition in this case with multiple interacting wall surfaces), such interference may divert the turtle from its intended route, forcing it to expend additional time correcting its course back to the correct path.

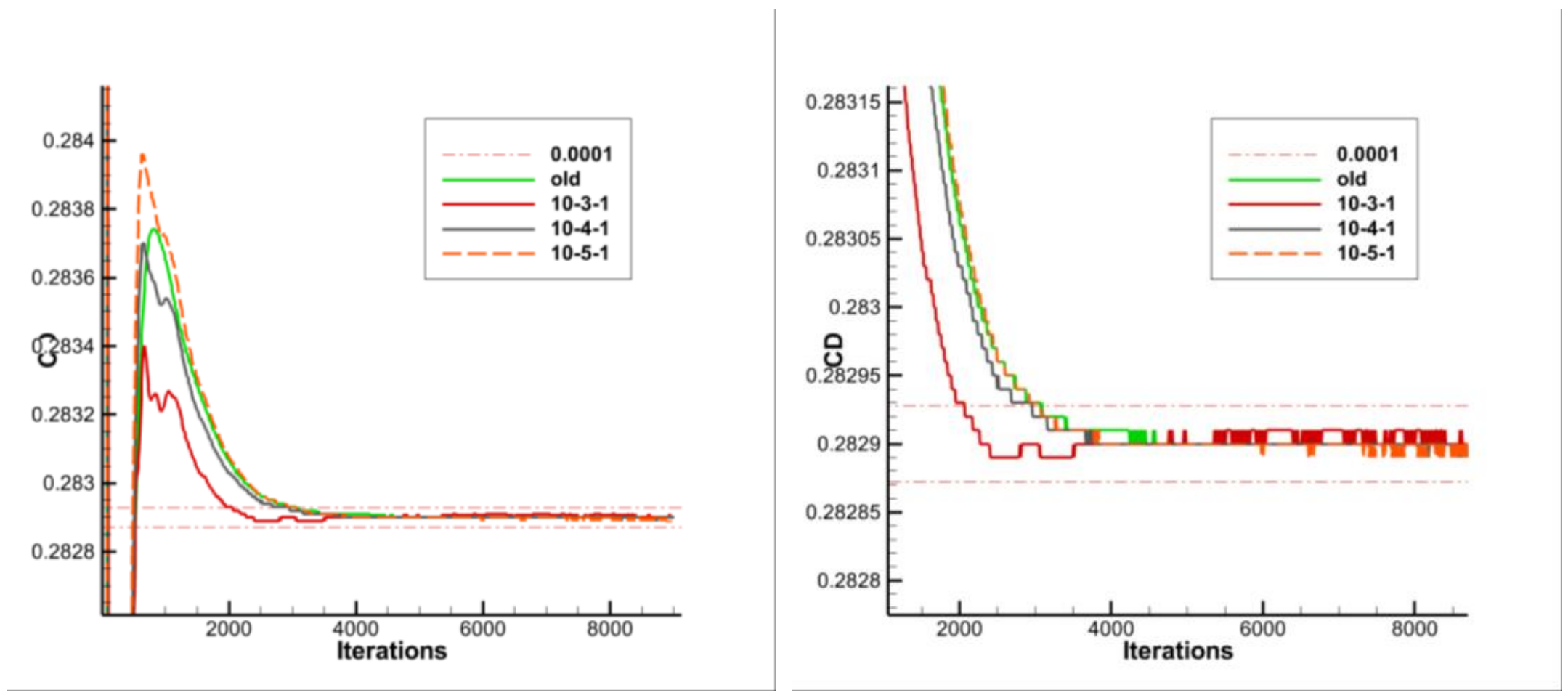

Figure 17 Comparison of Drag Coefficient Convergence Curves for Different Iterative Modes

Figure 18 presents the pressure contours and streamlines on the upper wing surface obtained by the original method and the “10-3-1” mode under this condition. It can be observed that no significant separation region exists on the upper surface and the flow field results from both methods are entirely consistent.

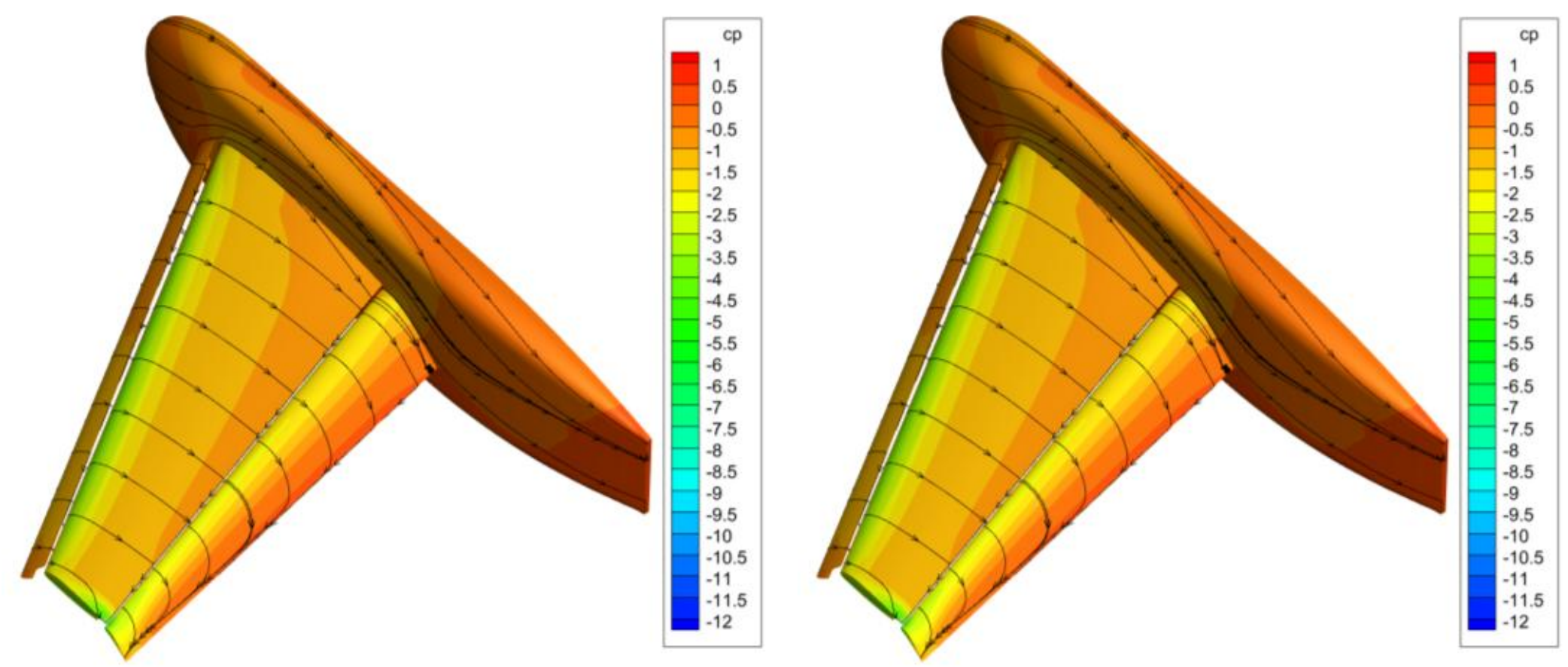

Figure 18 Upper surface Pressure Contours and Streamlines at AoA= 11.02° (Left: Original Method, Right: New Method)

Figure 19 compares the convergence histories of different computational modes at AoA = 24.40°, with an enlarged provided on the right. Under this high angle of attack condition, the comparison of convergence steps between different modes reveals a new trend again. The convergence speed of the “10-3-1” mode is significantly slower than that of the original method. However, as the number of inner iterations increases, the iteration steps gradually decreases and eventually surpasses that of the original mode in the “10-8-1” mode. When the number of inner iterations is further increased—even exceeding the boundary layer iteration steps—the advantage in convergence speed becomes more pronounced, reaching its maximum in the “10-14-1” mode. The behavior of inner iteration steps in this case exhibits an almost opposite trend compared to the previous state. Specifically, a higher number of inner iterations leads to faster convergence. This study suggests that under high angle of attack conditions, the increased flow parameter gradients in the inner region require more iterations to resolve adequately. In contrast, the previous state at a moderate angle of attack involved relatively mild flow variations in the inner field. Excessive inner iterations in that case caused overshooting of flow parameters, thereby slowing down the overall convergence.

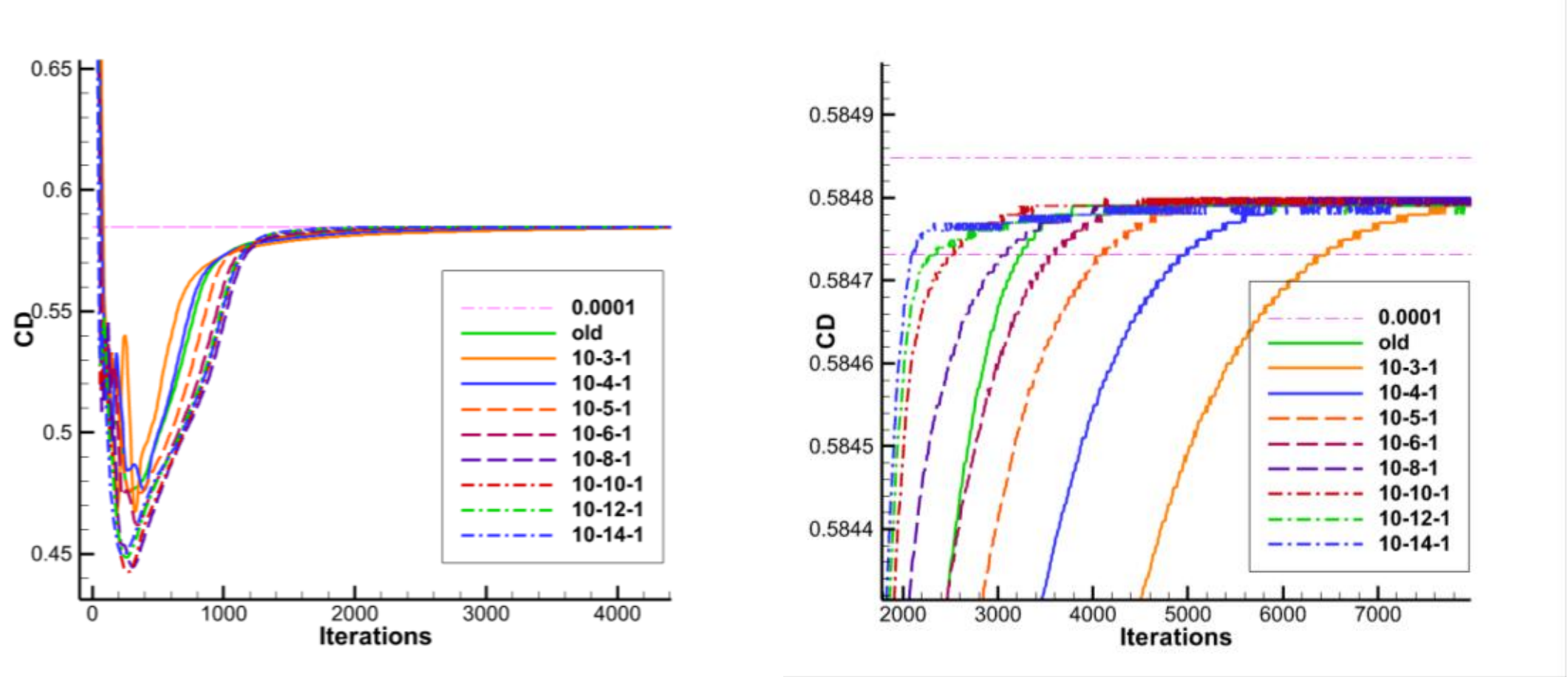

Figure 19 Comparison of Drag Coefficient Convergence Curves for Different Iterative Modes

Figure 20 presents the upper surface pressure contours and streamlines obtained by the original method and the “10-8-1” mode under this condition. It can be observed that the flow field results from both methods are entirely consistent.

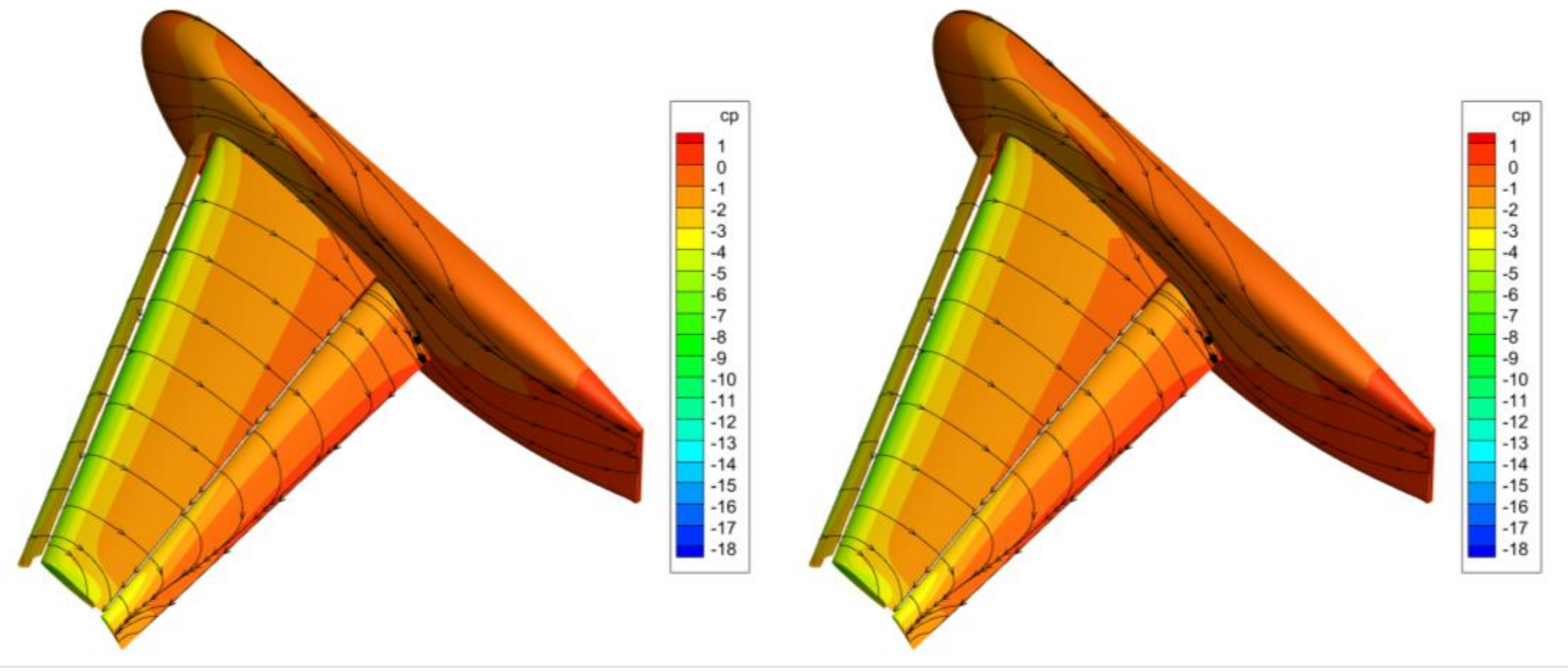

Figure 20 Upper surface pressure contours and streamlines at AoA = 24.40° (left: original method; right: new method).

Since an optimal unified computational mode could not be determined for this test case, a comparative study using different modes was conducted for all major flow conditions. Table 10 presents a comparison of the iteration steps and convergence time across various angles of attack and computational modes. For each angle of attack column, the first row indicates the number of iterations required for convergence (Nc) and the second row shows the corresponding convergence time. Values highlighted in red represent the fastest convergence iteration steps for that condition, while green highlights indicate the shortest convergence time. The rightmost column provides a comparison of the differences in the fastest iteration steps and convergence time relative to the original method. It can be observed that for angles of attack below 6.48°, the convergence efficiency among different modes shows little variation, with all iteration steps being comparable to those of the original method. Within the range of 11.02° to 20.02°, there almost consistently exists an optimal iterative mode that yields significantly fewer iterations than both the original method and other modes. For angles of attack exceeding 22.23°, the trend indicates that higher numbers of inner iterations result in faster convergence.

Table 10 Computational Efficiency Comparison of Different Modes for the Trap Wing

| AoA (°) | | Old | 10-3-1 | 10-4-1 | 10-5-1 | 10-6-1 | 10-7-1 | 10-8-1 | 10-12-1 | Iteration Difference |
|---|---|---|---|---|---|---|---|---|---|---|
| -3.45 | Nc | 24670 | 24460 | 23700 | 24160 | 24150 | / | / | / | 96.07 % |
| | Time(s) | 47736.5 | 25095.9 | 25554.5 | 27312.8 | 28678.1 | / | / | / | **52.57%** |
| 1.83 | Nc | 5980 | 6320 | 6100 | 6080 | 6140 | 6130 | / | / | 101.67% |
| | Time(s) | 11571.3 | 6484.3 | 6577.3 | 6873.4 | 7291.2 | 7628.8 | / | / | **56.04%** |
| 6.48 | Nc | 6610 | 6730 | 6760 | 6860 | 6700 | 6800 | 6780 | / | 101.36% |
| | Time(s) | 12790.3 | 6904.9 | 7288.9 | 7755.2 | 7956.2 | 8462.6 | 8824.1 | / | **53.99%** |
| 11.02 | Nc | 3720 | 2480 | 3790 | 3930 | 3930 | / | / | / | 66.67% |
| | Time(s) | 7198.2 | 2544.4 | 4086.5 | 4442.8 | 4666.8 | / | / | / | **35.35%** |
| 15.53 | Nc | 3190 | 2960 | 2850 | 3520 | 3480 | 3490 | 3500 | / | 89.34% |
| | Time(s) | 6172.6 | 3036.9 | 3073.0 | 3979.3 | 4132.5 | 4343.3 | 4555.2 | / | **49.20%** |
| 17.78 | Nc | 2220 | 4140 | 3010 | 2100 | 3400 | 3340 | 3140 | / | 94.59% |
| | Time(s) | 4295.7 | 4247.6 | 3245.5 | 2374.0 | 4037.5 | 4156.6 | 4086.7 | / | **55.27%** |
| 20.02 | Nc | 3030 | 5460 | 4140 | 3340 | 2590 | 2200 | 1990 | 2480 | 65.68% |
| | Time(s) | 5863.0 | 5601.9 | 4463.9 | 3775.8 | 3075.6 | 2737.9 | 2589.9 | 3769.6 | **44.17%** |
| 22.23 | Nc | 3270 | 6340 | 4900 | 3940 | 3210 | 2880 | 2600 | 2260 | 69.11% |
| | Time(s) | 6327.4 | 6504.8 | 5283.4 | 4454.2 | 3811.8 | 3584.1 | 3383.9 | 3435.2 | **53.48%** |
| 24.40 | Nc | 3910 | 7790 | 6010 | 4980 | 4340 | 3960 | 3720 | 2840 | 72.63% |
| | Time(s) | 7565.8 | 7992.5 | 6480.2 | 5629.8 | 5153.7 | 4928.2 | 4841.5 | 4316.8 | **57.06%** |

However, this study suggests that the seemingly divergent trends observed across different angle of attack ranges actually follow a consistent underlying principle: an optimal iterative mode exists for all conditions that minimizes the number of convergence steps. A preliminary discussion is provided below. First, based on the comparative results in this case, varying the number of outer field iterations (from 1 to higher values) has negligible impact on the convergence history across all tested conditions. Since the number of boundary layer iterations was fixed at 10, the optimal iterative mode for each condition is solely determined by the number of inner field iterations. Therefore, the following discussion focuses exclusively on the effect of inner field iterations on convergence behavior. Second, in the low angle-of-attack regime (below 6.48°), the convergence steps are notably high generally exceeding 6,000 iterations, and even surpassing 24,000 under negative angles of attack. This is primarily due to the presence of extensive flow separation in these conditions. The slow evolution of separation regions during the iteration process significantly delays convergence. In contrast, for angles of attack above 11.02°, boundary layer separation is markedly reduced, with only small separation bubbles observed on both the upper and lower surfaces. Figure 21 presents surface streamlines and skin friction contours near the wing-body junction at various angles of attack, clearly illustrating the variation of separation regions with angle of attack. This reduction in separation is the main reason for the sharp decrease in convergence steps beyond 11.02°. It is worth noting that the large separation on the lower surface observed at AoA= -3.45° (Figure 16) does not occur at other angles of attack, which explains why this condition exhibits the slowest convergence.

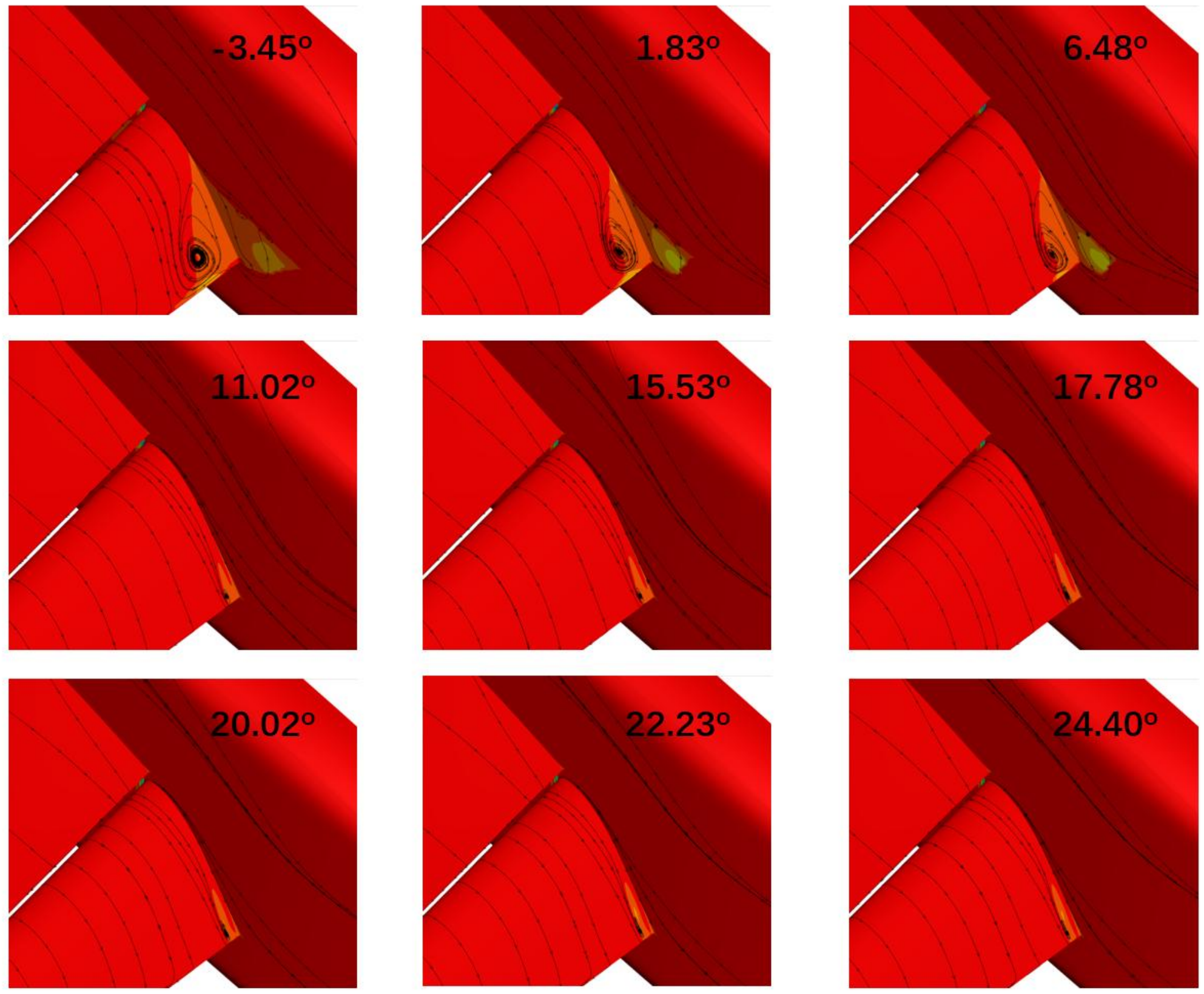

Figure 21 Surface streamlines at the wing-body junction for different angles of attack

Analysis across different angle of attack ranges:

1) As observed in the convergence history at AoA = -3.45° (Figure 15), the convergence curves of all modes exhibit a noticeable overshoot around 3,500 iterations, followed by a gradual decline to the converged value. Similar to previously discussed conditions, although the magnitude of this overshoot varies slightly depending on the computational mode, the convergence process requires approximately 20,000 iterations to enter the asymptotic convergence region. During this prolonged decline process, the convergence step advantage brought by the optimal inner field iterative mode "10-4-1" is almost entirely negated, failing to distinguish itself from other states. Other low angle of attack conditions exhibit similar convergence behavior, which explains why the convergence steps for these conditions appear very similar to those of the original mode.

2) For angles of attack beyond 11.02°, the virtual disappearance of separated regions leads to a significant reduction in convergence steps, averaging around 3,000 iterations. As shown in the convergence history at AoA=11.02° (Figure 17), the number of iterations required for the convergence curve to settle into the asymptotic region is substantially lower under these conditions. In such cases, the improvement of several hundred to even a thousand steps brought by the optimal inner field iterative mode becomes clearly distinguishable.

3) For angles of attack exceeding 22.23°, an optimal mode that maximizes convergence speed still exists. However, this mode may require a notably higher number of inner iterations than boundary layer iterations, reducing computational efficiency per step and potentially diminishing cost-effectiveness. Figure 22 further illustrates the impact of varying inner iteration counts on the convergence history at AoA = 24.40°. It can be seen that when the inner iteration count increases to 18–26, although these modes enter the convergence region earlier, their convergence curves differ significantly from those with fewer inner iterations (e.g., the “10-14-1” mode). These high inner iterative modes exhibit a distinct overshoot-and-decline pattern within the convergence region, with the overshoot magnitude increasing alongside the inner iteration count. This observation further supports the conclusion of this study: excessive inner iterations may cause overshooting of flow parameters and potentially delay global convergence. Nevertheless, due to the high drag coefficient under this condition, the convergence criterion of ±0.0001CD used in this study still accommodates the overshoot magnitude of these modes. Should a stricter convergence threshold be applied, the “10-14-1” mode would likely emerge as the optimal choice with the fewest iterations.

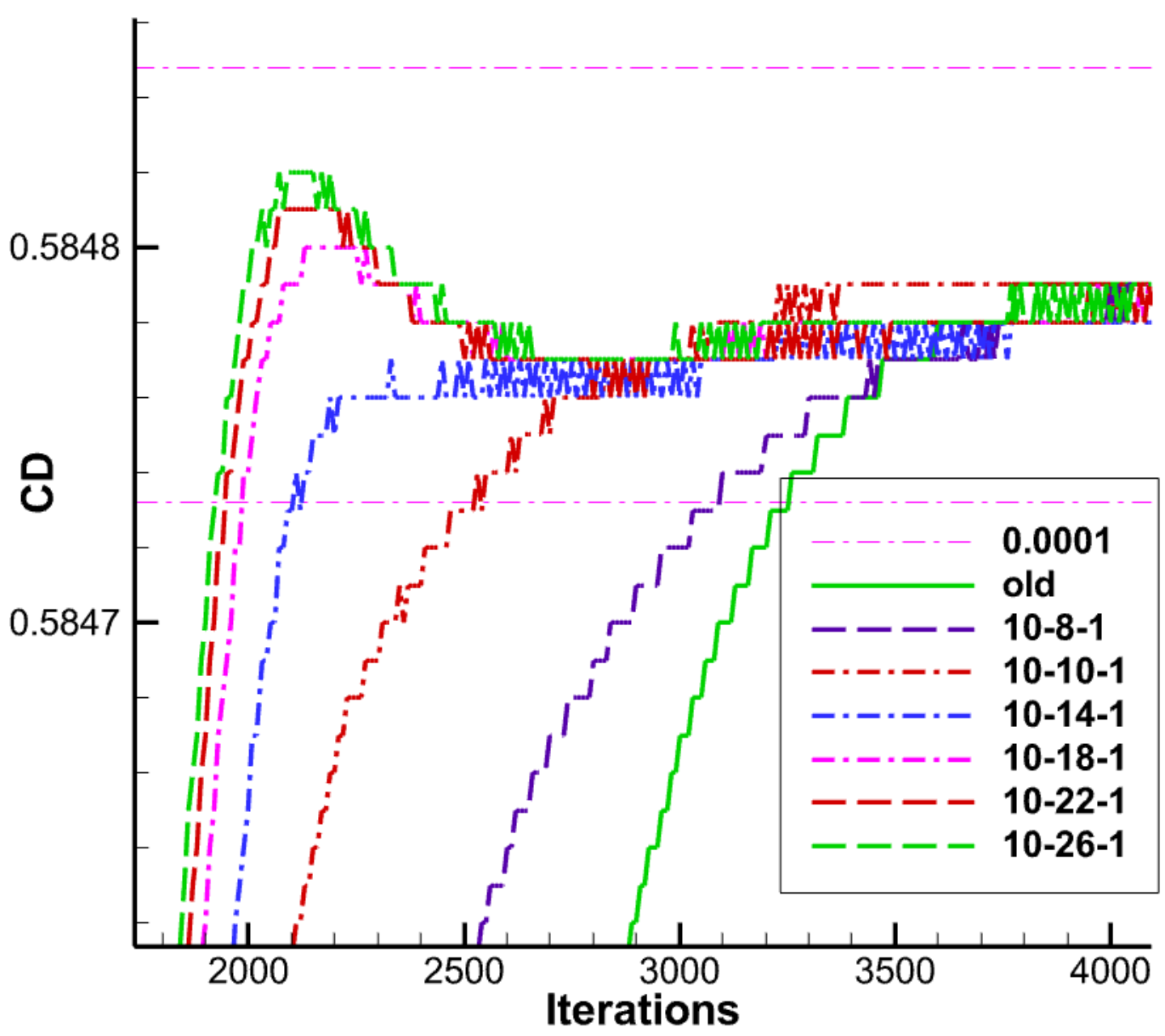

Figure 22 Comparison of convergence histories for different computational modes at AoA = 24.40°

However, the aforementioned analysis of the optimal iterative mode focuses solely on minimizing the number of convergence steps. When considering the computational time per iteration, not all modes with the fewest convergence steps prove to be practically optimal. As shown in Table 10, the actual optimal mode is “10-3-1” at low angles of attack. As the angle of attack increases, the optimal mode gradually shifts to “10-5-1” and “10-8-1”, and finally to “10-12-1” at the highest angle of attack. This trend indicates that higher angles of attack generally require more inner iterations, which aligns with the physical phenomenon of increasing flow parameter gradients in the inner region as the angle of attack grows. Furthermore, at AoA = 11.02°, 20.02°, and 22.23°, the fastest convergence requires less than 70% of the iterations needed by the original method. At AoA = 11.02°, the reduction in iteration count combined with shorter computational time per step results in total convergence time being only one-third of that of the original method. Averaging the computational time across all conditions for this test case, the new method achieves identical results using only 50.85% of the computational time required by the original method. For this highly complex configuration, the overall acceleration benefit of the new method slightly exceeds that observed in the two simpler test cases discussed earlier.

Figure 23 presents a comparison of the aerodynamic forces between the computational results obtained using the new method and the experimental data. It can be observed that across the range from negative to high angles of attack, the lift and drag coefficients calculated by the present method show excellent agreement with the experimental values. This demonstrates that the proposed hierarchical asynchronous iterative method also exhibits very good applicability for computationally complex configurations of this type.

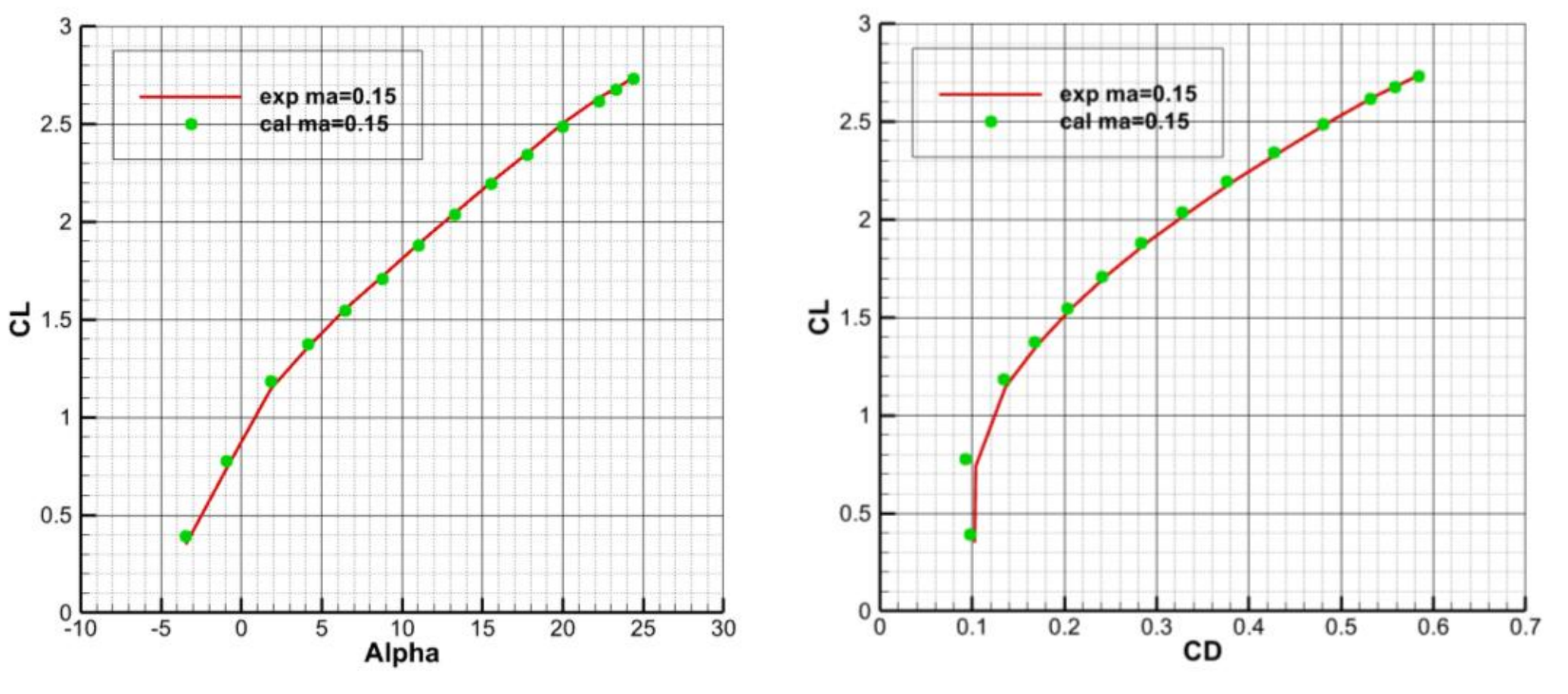

Figure 23 Comparison of experimental and computational aerodynamic force coefficients

## 5. Discussion and Recommendations

The computational results from the aforementioned test cases demonstrate the feasibility of the hierarchical iterative method. Building upon this foundation, a wider range of solution strategies can be derived based on hierarchical modes. Due to space limitations, this paper does not provide numerical validation for additional hierarchical solution strategies, but offers the following explanations and recommendations regarding the proposed method:

1) During the grid generation process in this study, the height of the boundary layer region is typically set to approximately 2–3 times the estimated turbulent boundary layer height based on the mean aerodynamic chord (MAC). If the MAC is relatively small compared to the overall vehicle length (as in the DLR-F6 case presented in this paper), this ratio may be appropriately increased to 3–4 times. Numerical simulation conducted in this work indicate that an insufficiently thick boundary layer region may lead to degraded convergence behavior in hierarchical iterative modes. Furthermore, the spatial extent of the interior field should generally encompass regions with significant flow gradients. In this study, the streamwise extension is typically set to 1/3–1/2 of the fuselage length, the spanwise extension to approximately half of the semi-span length, and the normal extension comparable to the other two directions.

2) In this study, the number of grid points within the boundary layer region was set between 37 and 41, ensuring that the number of points across the turbulent boundary layer height exceeds 30 to meet the grid density requirements for turbulence modeling.

3) The acceleration effect of the hierarchical iterative method proposed in this study depends on the grid density within the boundary layer. Sparse boundary layer grids, where the number of cells in this region accounts for a smaller proportion of the total grid, lead to better acceleration performance. The present computations primarily employ a low-Reynolds-number turbulence model with $y^+ \approx 1$, resulting in relatively high grid density in the boundary layer. If a high-Reynolds-number turbulence model, such as the k-ε model with wall functions, is used for simulation, the height of the first grid cells can be significantly increased to $y^+ > 30$. This would substantially reduce the number of grid points in the boundary layer, further enhancing the advantage of the hierarchical iterative mode.

4) The Navier-Stokes equations admit a thin-layer approximation formulation (TLNS). When this approximation is employed, the streamwise viscous diffusion effects within the boundary layer are considered negligible compared to those in the normal direction. Consequently, only the second-order viscous derivatives in the normal direction are retained in the diffusion term, while those in other directions are neglected, significantly reducing the computational cost of the diffusion term. Within the framework of the hierarchical iterative method, the flow field is explicitly partitioned into 3 layers, enabling the use of different governing equations tailored to the flow characteristics of each region. Numerical experiments in this study demonstrate that in most cases, employing the TLNS equations in the boundary layer region while retaining the full NS equations in the inner and outer fields yields aerodynamic results that differ by approximately 1‰ from those obtained using the full NS equations throughout. Furthermore, this hybrid approach can provide an additional efficiency gain of 2–3%.

5) Different CFL numbers can be selected for flow field iterations in different layers to accelerate convergence. However, when an adaptive local time-stepping technique is employed, the benefits of this strategy become less pronounced.

6) Different control parameters can also be applied to the multigrid method across different layers, such as using distinct multigrid relaxation factors for each layer. During the prolongation process from coarse to fine grids in multigrid techniques, the interpolated flow variable corrections are multiplied by a relaxation factor to balance convergence speed and computational robustness. This factor typically ranges between 0.5 and 0.8, and in practice, different relaxation factors may be selected for different variables. Excessively large relaxation factors may lead to computational divergence in regions with high velocity gradients within the boundary layer due to drastic changes in flow variables—a phenomenon particularly pronounced in complex geometries. Conversely, overly small relaxation factors reduce the efficiency of multigrid acceleration. With the hierarchical iteration framework, users can strategically assign different relaxation factors to specific grid layers: for instance, employing slightly smaller factors in the boundary layer to enhance robustness, while using larger factors in the inner and outer fields to reduce iteration counts. Furthermore, dynamic relaxation factors can be implemented across layers at different stages of the convergence process to better consider convergence speed and stability. Additionally, the multigrid inner iteration cycle in this study was set to 10 steps. If fewer steps are desired-e.g., 5 steps-users may construct a multi-level cycle such as “5-2-1”, or even combine two or more multigrid cycles into a single multi-layer iterative loop to achieve higher acceleration benefits.

7) The implementation strategy of turbulence models can also be tailored to different grid layers. For instance, the commonly used Detached Eddy Simulation (DES) method is a hybrid RANS/LES model that relies on wall distance to determine whether the RANS or LES approach should be applied in a specific region. With the hierarchical mode, users can more conveniently control the application areas of different models—for example, enforcing the use of a specific model in certain grid layers to achieve improved simulation results.

8) The numerical examples in this paper are based on the structured grid method, but the proposed method is equally applicable to the unstructured grid method. The specific implementation approach is provided here. After generating an unstructured grid using conventional methods, users need to add one additional step: partitioning the grid into different layers. For the boundary layer region, users can specify a boundary layer height and define all grid cells whose wall distance is less than this height as boundary layer cells. Since the wall distance for all grid cells must be computed before solving the turbulence model anyway, this step does not introduce any additional workload. For the inner field, users can define a suitable spatial region—such as the cuboid region used in this paper or other spherical regions—and define all grid cells within this region that do not belong to the boundary layer as inner field cells. The remaining spatial cells are then defined as outer field grid cells. After completing the layer partitioning, the same strategy as described earlier is adopted during parallel decomposition to ensure that the number of grid cells in each layer on each computing core is approximately the same, thereby achieving optimal parallel efficiency. Overall, since unstructured grids do not possess the concepts of spatial topology or grid blocks, the layering implementation process is simpler and more flexible.

9) For different grid layers, more distinct control parameters or even numerical solution methods can be applied. Since the computational grid is forcedly partitioned into layers according to its flow characteristics, users can conduct further research on iteration patterns for specific flow phenomena when solving particular problems, thereby obtaining an iteration strategy with optimal efficiency.

10) For complex flows, how to quickly determine the optimal iteration pattern is a problem that requires focused investigation. Based on current numerical experimental results, the optimal pattern for relatively simple flows is essentially fixed. For complex flows, as the angle of attack increases, a preliminary trend emerges that the number of inner-field iterations needs to be correspondingly increased. However, whether this trend is universally applicable, and how to obtain quantifiable criteria to minimize iteration time, both require extensive follow-up research. Of course, even if a fixed iteration pattern is adopted for flow field computation, although there is a slight gap compared with the optimal pattern, a good acceleration effect can still be achieved. It is our hope that, taking this paper as a starting point, more researchers will be attracted to conduct in-depth studies on the hierarchical iteration framework, thereby enabling this method—which is simple to implement yet yields significant benefits—to achieve wider application.

## 6. Conclusions

Through validation with three typical test cases covering different speed regimes, the proposed method achieved identical computational results while consuming only 53.2% of the computation time required by the conventional iterative approach on average, without significantly increasing manual effort. Based on the numerical experiments conducted in this study, the following conclusions can be drawn:

1) Within the hierarchical iteration framework proposed in this paper, the convergence results of the new method are unaffected by the number of iteration steps in each layer. While varying the iteration steps across different regions influences convergence efficiency, it does not alter the final converged solution.

2) In computational states with relatively simple flow field structures, where there is no significant interaction between multiple components or flow features, the number of iterations required for convergence is nearly identical between the proposed method and the conventional approach. In such cases, the benefit of the new method primarily stems from the reduction in computation time per iteration.

3) In computational states with complex flow field structures, the traditional synchronous iterative mode is not necessarily the fastest. Using a uniform number of iteration steps across all regions may lead to overshoots in flow parameters in certain areas, thereby increasing the total number of iterations required for convergence. In contrast, the new method almost always enables an optimal iterative mode that minimizes the number of convergence steps—often significantly outperforming the conventional approach. Under these conditions, the combined effect of reduced time per iteration and fewer convergence steps results in substantially improved computational efficiency.

4) In the transonic regime, as demonstrated in the test cases, it is necessary to appropriately increase the number of iterations in the outer field to achieve better convergence. However, for supersonic and low-speed flows, one iteration (or even fewer) in the outer field is sufficient.

5) For steady or quasi-steady simulations, rational application of the hierarchical iterative mode can greatly reduce computational resource consumption. Although unsteady cases require further validation, it is anticipated that similar acceleration effects can be achieved in unsteady simulations using the dual-time-step method, wherein the hierarchical iteration strategy is applied within the pseudo-time iterations.

6) The proposed hierarchical asynchronous iterative methodology is not limited to CFD applications. It can also be applied to numerical solutions of spatial problems in other fields that employ finite difference methods or finite volume methods, such as electromagnetics, astrophysics, and petroleum reservoir simulation.